\documentclass[reprint,superscriptaddress,amsmath,amssymb,aps,prx,floatfix]{revtex4-2}
\usepackage{pdfpages,pgffor,graphicx,dcolumn,bm,xr,physics,chemformula,siunitx,color,soul,float,appendix, subfigure, booktabs,array, multirow, pgfplots}
\pgfplotsset{compat=1.18}
\usepgfplotslibrary{external}
\usepackage[dvipsnames]{xcolor}

\def\bk{{\bf k}}
\def\br{{\bf r}}

\def\bq{{\bf q}}

\def\bu{{\bf u}}

\def\bR{{\bf R}}
\def\ba{{\bf a}}
\def\bA{{\bf A}}

\def\bb{{\bf b}}
\def\bB{{\bf B}}

\def\bg{{\bf g}}
\def\bG{{\bf G}}

\def\a{\alpha}

\def\w{\omega}
\def\k{\kappa}

\def\ve{\varepsilon}
\def\btau{\boldsymbol{\tau}}

\def\D{\partial}
\def\d{\delta}
\def\<{\langle}
\def\>{\rangle}

\def\had{\hat{a}^\dagger}
\def\ha{\hat{a}^{\vphantom{\dagger}}} 

\def\hcd{\hat{c}^\dagger}
\def\hc{\hat{c}^{\vphantom{\dagger}}} 

\def\hH{{\hat{H}}}

\def\dtau{\Delta{\tau}}

\def\gone{g^{(1)}}
\def\gtwo{g^{(2)}}

\newcommand{\red}[1]{\textcolor{black}{#1}}

\makeatletter
\AtBeginDocument{\let\LS@rot\@undefined}
\makeatother

\begin{document}

\raggedbottom

\title{Nonlinear electron-phonon interactions from first principles}

\author{Zhenbang Dai}
\affiliation{Oden Institute for Computational Engineering and Sciences, The University of Texas at Austin, Austin, Texas 78712, USA}
\affiliation{Department of Physics, The University of Texas at Austin, Austin, Texas 78712, USA}
\author{Feliciano Giustino}%
\email{fgiustino@oden.utexas.edu}
\affiliation{Oden Institute for Computational Engineering and Sciences, The University of Texas at Austin, Austin, Texas 78712, USA}
\affiliation{Department of Physics, The University of Texas at Austin, Austin, Texas 78712, USA}
        
\date{\today}

\begin{abstract}
Electron-phonon interactions underpin a variety of phenomena, ranging from transport and superconductivity to polarons and ultrafast carrier dynamics. 
Despite being one of the most intensely studied subjects in condensed matter physics, research on electron-phonon physics mostly focused on linear, first-order couplings. Second- and higher-order nonlinear couplings are commonly ignored because their calculations are too demanding and we lack computational frameworks that can provide both diagonal and off-diagonal coupling matrix elements.  
In this work, we report a theory and computational method for computing nonlinear electron-phonon interactions of any order in real materials. 
Our approach combines the advantages of unit-cell calculations of electron wavefunctions and supercell calculations of phonon perturbations, is systematically improvable, can be used with either semilocal or nonlocal exchange-correlation functionals, and is amenable to Wannier-Fourier interpolation. 
As a first proof of concept, we illustrate this method by computing second-order electron-phonon coupling matrix elements in diamond, lithium fluoride, and graphite as representative nonpolar semiconductors, polar semiconductors, and metals, respectively. Furthermore, we generalize the \textit{ab initio} polaron equations to second-order electron-phonon couplings, and we show that second-order couplings are essential to achieve quantitative accuracy in polaron formation energies and hopping barriers. 
The present methodology will find application in the study of all properties and phenomena that are currently being investigated within the linear electron-phonon coupling approximation, from phonon-mediated superconductivity to excited-states dynamics, both in harmonic and anharmonic systems.

\end{abstract}
\maketitle

\section{Introduction}
\label{Sec:Intro}

Electron-phonon interactions have been at the center stage of condensed matter and materials physics since the establishment of this concept nearly a century ago~\cite{Bloch_1929}. They are responsible for countless phenomena, including phonon-assisted optical absorption~\cite{Noffsinger_Cohen_2012, Tiwari_Giustino_2024}, phonon-assisted shift and ballistic currents~\cite{Dai_Rappe_2021, Hu_Chang_2024}, phonon-limited electrical transport~\cite{Ponce_Giustino_2018, Zhou_Bernardi_2016}, pairing in conventional superconductivity~\cite{Bardeen_Schrieffer_1957, Margine_Giustino_2013}, satellites and kinks in photoemission spectroscopy~\cite{Verdi_Giustino_2017, Zhou_Bernardi_2019}, Kohn anomalies~\cite{Kohn_1959, Lazzeri_Mauri_2006}, and polaron physics~\cite{Frohlich_1954, Feynman_1955, Lee_Pines_1953}. 
The study of electron-phonon interactions has significantly advanced our understanding of these fundamental phenomena and enabled predictive calculations of related functional properties~\cite{Giustino_2017}. 

Given the ubiquity of electron-phonon interactions in condensed matter, it is somewhat surprising that most research in this area rests on the approximation of linear electron-phonon couplings. In this approximation, the change of electron energies resulting from the vibration of a given atom is proportional to amplitude of that vibration. However, it has been known for at least half a century that the energy change resulting from the quadratic term is equally important in the study of temperature-dependent band structures~\cite{Allen_Heine_1976}. In fact, this term is responsible for the Debye-Waller self-energy, which is comparable in magnitude and opposite in sign to the Fan-Migdal self-energy resulting from linear couplings~\cite{Giustino_Cohen_2010, Ponce_Park_2025}. Furthermore, it has already been pointed out that the inclusion of quadratic terms may impact, among others, calculations of phonon-assisted optical spectra~\cite{Tiwari_Giustino_2024, Dai_Rappe_2021} and electrical transport~\cite{Houtput_Franchini_2026}. 
It is worth mentioning that ``electron-multiphonon'' couplings have been investigated in the case of electrical transport~\cite{Lee_Bernardi_2020} and superconductivity~\cite{Mishra_Margine_2025}, but these terms correspond to sequential electron-one-phonon processes.

More recently, work on systems such as high-pressure hydride superconductors~\cite{Errea_Flores-Livas_2020, Hou_Errea_2021, Belli_Errea_2021, Bianco_Errea_2023} and halide perovskites~\cite{Yaffe_Brus_2017, Zacharias_Even_2023a, Zacharias_Even_2023b} highlighted the importance of anharmonic effects, which correspond to beyond-quadratic terms in the expansion of the Born-Oppenheimer potential energy surface near the ground-state structure. These studies naturally raise the question on whether higher-order terms in the electron-phonon couplings should similarly be included. Furthermore, in the context of effective Hamiltonian approaches, it was found that the inclusion of nonlinear electron-phonon couplings in the Migdal-Eliashberg theory may increase the superconducting critical temperature~\cite{Houtput_Tempere_2025, Zappacosta_Tempere_2025}. Similarly, nonlinear electron-phonon couplings are thought to be important in the context of polaron physics~\cite{Ragni_Mishchenko_2023, Ragni_Mishchenko_2025}. For example, we recently found that polaron energetics computed from linear electron-phonon Hamiltonians can differ significantly from the result of density-functional theory (DFT) supercell calculations that include nonlinear couplings~\cite{Dai_Giustino_2026}.

Despite the increasing evidence for the importance of nonlinear electron-phonon couplings, \textit{ab initio} calculations of the corresponding matrix elements are scarce, and calculations of the off-diagonal matrix elements are non-existent to the best of our knowledge. 
The scarcity of work in this area is due to the fact that obtaining nonlinear electron-phonon coupling matrix elements is theoretically and computationally challenging. Within density functional perturbation theory (DFPT), the calculation of nonlinear couplings would require the extension of the first-order Sternheimer equation into higher-order equations~\cite{Baroni_Gianozzi_2001, Giustino_2017}; this is possible in principle but technically challenging, and has not been achieved thus far.
Alternatively, it is possible to compute these matrix elements via finite differences using supercells, as shown for example in the case of molecular systems~\cite{Gonze_Cote_2011}; or by expressing the sum of diagonal second-order matrix elements via sums of first-order matrix elements~\cite{Giustino_Cohen_2010, Allen_Heine_1976, Allen_Cardona_1981}, or using the commutator of the position operator and the first-order coupling potential~\cite{Lihm_Park_2020}. While these approaches address some of the limitations of linear electron-phonon theories, their applications to solids are all based on the so-called ``rigid-ion approximation'' (RIA), whose reliability remains untested, and they only provide certain diagonal matrix elements rather than the full coupling matrix.

Here, we fill this gap by developing a first-principles framework to compute nonlinear electron-phonon couplings that overcomes the limitation of existing approaches, is not restricted to the RIA, and enables calculations of both diagonal and off-diagonal matrix elements of any order. As a first demonstration of this method, we focus on second-order electron-phonon coupling matrix elements. Our key findinds are: (i) a computationally tractable expression for the second-order electron-phonon coupling matrix elements that is amenable to first principles calculations, and (ii) the generalization of the \textit{ab initio} polaron equations to include second-order electron-phonon couplings.

The article is organized as follows: in Sec.~\ref{Sec:Theory} we outline the theoretical framework. In particular, in  Sec.~\ref{Sec:Derivations} we derive the expressions for the nonlinear electron-phonon coupling matrix elements, and we analyze in detail the second-order couplings;
in Sec.~\ref{Sec:RIA} we review the RIA, and in Sec.~\ref{Sec:RIA_better} we discuss systematic improvements upon this approximation;
Sec.~\ref{Sec:Polaroneq} presents the generalization of the \textit{ab initio} polaron equations to second-order couplings;
and in Sec.~\ref{Sec:Wannier} we outline the generalization of electron-phonon Wannier interpolation to the case of second-order coefficients. 
Section~\ref{Sec:numerical} discusses the pratical implementation of this method: in Sec.~\ref{Sec:Computation} we outline the workflow and the computational setup used in this work; in Sec.~\ref{Sec:Conv_dtau} we apply this method to compute the second-order matrix elements for diamond, lithium fluoride (LiF), and graphite; in Sec.~\ref{Sec:RIA_bad} we analyze the accuracy of the RIA and its improvements.
Section~\ref{Sec:Application} is devoted to polarons: in Sec.~\ref{Sec:Polaroneq_sol} we solve the second-order polaron equations to obtain the formation energy of the small hole polaron in LiF; and in Sec.~\ref{Sec:Hopping} we investigate the corresponding polaron hopping barriers and mobilities.
In Sec.~\ref{Sec:Conclusion} we summarize our main findings and discuss potential applications of this method and further extensions.
For completeness, in App.~\ref{App:band_conv}, we investigate the convergence of the polaron equations with respect to the the number of bands.

\section{Theoretical framework}
\label{Sec:Theory}

\subsection{Procedure for calculating nonlinear electron-phonon couplings}
\label{Sec:Derivations}

In this section we outline the procedure for calculating nonlinear electron-phonon couplings. In particular, we derive the expressions for the second-order electron-phonon coupling matrix elements. At the end of this section, we show how the same expressions can straightforwardly be generalized to higher-order couplings.

Our aim is to calculate the complete second-order electron-phonon coupling matrix elements $g^{(2)}_{mn\nu\nu'}(\bk, \bq, \bq')$, as defined in Ref.~\citenum{Giustino_2017}: 
\begin{align}
    g^{(2)}_{mn\nu\nu'}(\bk,\bq,\bq')
    =&
    \frac{1}{2}
    \sum_{\substack{\k \a p\\ \k' \a' p'}}
    \sqrt{ \frac{\hbar} {2M_\k \w_{\bq\nu}}}
    e_{\k \a,\nu}(\bq)
    e^{i\bq\cdot\bR_p}
    \nonumber \\
    &\hspace{10pt}\times
    \sqrt{ \frac{\hbar} {2M_{\k'} \w_{\bq'\nu'}}}
    e_{\k'\a',\nu'}(\bq')
    e^{i\bq'\cdot\bR_{p'}}
    \nonumber \\
    &\hspace{10pt}\times \< \psi_{m\bk+\bq+\bq'}| \frac{\partial^2 V_{\rm KS}}{\partial \tau_{\k\a p} \partial \tau_{\k'\a' p'}} |\psi_{n\bk}\>_{\rm sc}.
    \label{Eq:2ndeph_def}
\end{align}
In this expression, $\psi_{n\bk}$ denotes the Kohn-Sham wavefunction for band $n$ and wavevector $\bk$; this wavefunction is normalized in the Born-von-K\'{a}rm\'{a}n (BvK) supercell. $e_{\k \a,\nu}(\bq)$ is a vibrational eigenmode with branch index $\nu$ and wavevector $\bq$, with frequency $\w_{\bq\nu}$; $\k$ and $\a$ are the atom label in the unit cell and the Cartesian direction, respectively. $M_\k$ represents the mass of atom $\k$. $\bR_p=n_1 \ba_1 + n_2 \ba_2 + n_3 \ba_3$ is a vector of the direct lattice, with $\ba_\a$ denoting primitive lattice vectors and the $n_\a$ are integers. $V_{\rm KS}$ is the Kohn-Sham potential, and the bra-ket $\< \cdots \>_{\rm sc}$ indicates the integral over the BvK supercell.

The second-order matrix elements appear in the second-order electron-phonon Hamiltonian through the term~\cite{Giustino_2017}:
\begin{eqnarray}\label{Eq:H2}
 \hH^{(2)}_{\rm ep} &=&\frac{1}{N_p}\!\!
     \sum_{\substack{\bk,\bq,\bq'\\m n \nu\nu'}} \!g^{(2)}_{mn\nu\nu'}(\bk,\bq,\bq')\,
     \hcd_{m\bk+\bq+\bq'} \hc_{n\bk} \nonumber \\
&\times&
   (\ha_{\bq\nu}+\had_{-\bq\nu}) (\ha_{\bq'\nu'}+\had_{-\bq'\nu'}),    
\end{eqnarray}
where $N_p$ is the number of unit cells in the BvK supercell, and the wavectors in the sum belong to a uniform Brillouin zone grid with $N_p$ points. The $\hc_{n\bk}/\hcd_{n\bk}$ operators in this expression are the fermionic operators, and the $\ha_{\bq\nu}/\had_{\bq\nu}$ are the bosonic operators. The physical meaning of $g^{(2)}_{mn\nu\nu'}(\bk,\bq,\bq')$ is the probability amplitude for an electron in state $\psi_{n\bk}$ to transition into state $\psi_{m\bk+\bq+\bq'}$ by absorbing or emitting a phonon with branch $\nu$ and wavevector $\bq$ and simultaneously absorbing or emitting a phonon with branch $\nu'$ and wavevector $\bq'$. 
By taking the expectation value of the Hamiltonian in Eq.~\eqref{Eq:H2} over the many-body state with one electron in $|\psi_{n\bk}\>$, one obtains immediately the Debye-Waller correction to the electron quasiparticle energy due to zero-point fluctuations~\cite{Giustino_2017}:
\begin{equation}
    \label{Eq:dw_selfe}
    \Delta \varepsilon_{n\bk}^{\rm DW} 
    = \< \psi_{n\bk}|  \hH^{(2)}_{\rm ep} |\psi_{n\bk}\>
    = \frac{1}{N_p}\sum_{\bq\nu} g^{(2)}_{nn\nu\nu}(\bk,\bq,-\bq).
\end{equation}
The corresponding expression at finite temperature is obtained by carrying out a standard thermal average and can be found in Ref.~\citenum{Giustino_2017}. This Debye-Waller energy shift is one of two ingredients (the other being the Fan-Midgal self-energy) that are needed to compute temperature-dependent band structures within the Allen-Heine theory~\cite{Allen_Heine_1976,Allen_Cardona_1981,Giustino_Cohen_2010,Ponce_Gonze_2015,Ponce_Park_2025}.

Of the ingredients appearing in Eq.~\eqref{Eq:2ndeph_def}, the Kohn-Sham wavefunctions are typically obtained from DFT, and the vibrational eigenmodes and eigenfrequencies are routinely computed using DFPT~\cite{Baroni_Gianozzi_2001} or via finite differences~\cite{Togo_Tanaka_2023}. Therefore, the only challenging part of Eq.~\eqref{Eq:2ndeph_def} is the evaluation of the BvK integral in the last line, 
 \begin{equation}\label{Eq:d2vdtau2}
     \< \psi_{m\bk+\bq+\bq'}| \frac{\partial^2 V_{\rm KS}}{\partial \tau_{\k\a p} \partial \tau_{\k'\a' p'}} |\psi_{n\bk}\>_{\rm sc}~.
 \end{equation}
Direct evaluation of Eq.~\eqref{Eq:d2vdtau2} via finite differences is impractical, because it would require the computation of $(3N_{\rm at} N_p)^2$ second-order derivatives, where $N_{\rm at}$ is the number of atoms in the unit cell, as well a large number of Kohn-Sham wavefunctions in the BvK supercell. Here, we overcome these obstacles by exploiting (i) the lattice translational symmetry for the atomic positions, and (ii) the Bloch theorem for the electron wavefunctions. We begin by using the standard strategy employed for computing the dynamical matrix~\cite{Baroni_Gianozzi_2001} and electron-phonon matrix elements in the Wannier representation~\cite{Giustino_Louie_2007}, and perform a translation of the integration variable so that one of the atoms falls within the unit cell with $\bR_{p'}=0$ (the ``reference'' cell):
 \begin{align}
    \label{Eq:d2vdtau2_shift}
    & \hspace{-10pt}  \< \psi_{m\bk+\bq+\bq'}| \frac{\partial^2 V_{\rm KS}}{\partial \tau_{\k\a p} \partial \tau_{\k'\a' p'}} |\psi_{n\bk}\>_{\rm sc}
    \nonumber \\
    =&
    \int_{\rm sc}\!\!\!d\br\, \psi^*_{m\bk+\bq+\bq'}(\br+\bR_{p'}) 
    \frac{\D^2 V_{\rm KS}(\br+\bR_{p'})}{\D \tau_{\k \a p} \D \tau_{\k' \a' p'}}
    \psi_{n\bk}(\br+\bR_{p'})      
    \nonumber \\
    =&e^{-i(\bq+\bq')\cdot \bR_{p'}}
    \!\!\int_{\rm sc}\!\!\!d\br\, \psi^*_{m\bk+\bq+\bq'}(\br) 
    \frac{\D^2 V_{\rm KS}(\br)}{\D \tau_{\k \a, p-p'} \D \tau_{\k' \a' 0}}
    \psi_{n\bk}(\br),          
\end{align}
where $\bR_{p-p'}$ is a short for $\bR_p-\bR_{p'}$. Upon using this expression inside Eq.~\eqref{Eq:2ndeph_def} and relabeling $p-p'$ as $p$, we obtain:
\begin{eqnarray}
    &&g^{(2)}_{mn\nu\nu'}(\bk,\bq,\bq') = 
    \frac{1}{2} N_p\!\!\!
    \sum_{\k \a, \k' \a'}
    \sqrt{ \frac{\hbar} {2M_\k \w_{\bq\nu}}}\sqrt{ \frac{\hbar} {2M_{\k'} \w_{\bq'\nu'}}}
    \nonumber \\ && \hspace{10pt}\times\, 
    e_{\k \a,\nu}(\bq)e_{\k'\a',\nu'}(\bq')
    \nonumber \\
    &&\hspace{10pt}\times  \sum_p e^{i\bq\cdot\bR_p} \< \psi_{m\bk+\bq+\bq'}| \frac{\partial^2 V_{\rm KS}}{\partial \tau_{\k\a p} \partial \tau_{\k'\a' 0}} |\psi_{n\bk}\>_{\rm sc}.
    \label{Eq:2ndeph_simple}    
\end{eqnarray}
This expression implies that one only needs to evaluate $(3N_{\rm at})^2 N_p$ derivatives, with one of the atoms lying within the reference unit cell. One could also exploit crystal symmetry operations to further reduce the number of inequivalent derivatives, but we leave this refinement to future work.
In the following, we focus on implementations based on norm-conserving pseudopotentials, so that the additional complexities arising from the derivatives of the atomic projectors in the PAW method~\cite{Blochl_1994} can be avoided~\cite{Chaput_Tanaka_2019}. 

Starting from Eq.~\eqref{Eq:2ndeph_simple}, our strategy consists of (i) evaluating the second derivatives via their finite-difference expressions in supercell calculations, and (ii) using the electron wavefunctions from \textit{undistorted, unit cell calculations} in the integral. For example, in the cases when
the combined indices $(\k\a p)$ and $(k'\a' 0)$ differ, we evaluate the mixed derivatives using central finite differences: 
 \begin{eqnarray}
    \label{Eq:fd_mixed}
    &&\< \psi_{m\bk+\bq+\bq'}| \frac{\partial^2 V_{\rm KS}}{\partial \tau_{\k\a p} \partial \tau_{\k'\a' 0}} |\psi_{n\bk}\>_{\rm sc} \nonumber \\ && =
    \frac{1}{4\d^2}\Big[\<\psi_{m\bk+\bq+\bq'}|V_{\rm KS}(\tau^0_{\k \a p}+\d,\tau^0_{\k' \a' 0}+\d)|\psi_{n\bk}\>_{\rm sc}
    \nonumber \\ &&\hspace{25pt}+
    \<\psi_{m\bk+\bq+\bq'}|V_{\rm KS}(\tau^0_{\k \a p}-\d,\tau^0_{\k' \a' 0}-\d)|\psi_{n\bk}\>_{\rm sc}
    \nonumber \\ &&\hspace{25pt}-
    \<\psi_{m\bk+\bq+\bq'}|V_{\rm KS}(\tau^0_{\k \a p}+\d,\tau^0_{\k' \a' 0}-\d)|\psi_{n\bk}\>_{\rm sc}
    \nonumber \\ &&\hspace{25pt}-
    \<\psi_{m\bk+\bq+\bq'}|V_{\rm KS}(\tau^0_{\k \a p}-\d,\tau^0_{\k' \a' 0}+\d)|\psi_{n\bk}\>_{\rm sc}\Big] \hspace{10pt}\nonumber \\ && \hspace{25pt}+ \mathcal{O}(\d^2)~,
\end{eqnarray}
where the notation $V_{\rm KS}(\tau^0_{\k \a p}+\d,\tau^0_{\k' \a' 0}+\eta)$ is a short for the Kohn-Sham potential of a supercell where all atoms are in their equilibrium positions, except for atoms $(\k p)$ and $(\k' 0)$ which are displaced by $\d$ and $\eta$ along the Cartesian directions $\a$ and $\a'$, respectively. The corresponding expression for the case where the combined indices $(\k \a p)$ and $(\k' \a' 0)$ coincide reads:
 \begin{eqnarray}
    \label{Eq:fd_diag}
    &&\hspace{-20pt}\< \psi_{m\bk+\bq+\bq'}| \frac{\partial^2 V_{\rm KS}}{\partial \tau_{\k\a 0}^2} |\psi_{n\bk}\>_{\rm sc} \nonumber \\ && =
    \frac{1}{\d^2}\Big[\hspace{10pt}\<\psi_{m\bk+\bq+\bq'}|V_{\rm KS}(\tau^0_{\k \a 0}+\d)|\psi_{n\bk}\>_{\rm sc}
    \nonumber \\ &&\hspace{30pt}+
    \<\psi_{m\bk+\bq+\bq'}|V_{\rm KS}(\tau^0_{\k \a 0}-\d)|\psi_{n\bk}\>_{\rm sc}
    \nonumber \\ &&\hspace{25pt}-2
    \<\psi_{m\bk+\bq+\bq'}|V_{\rm KS}|\psi_{n\bk}\>_{\rm sc}\Big]+ \mathcal{O}(\d^2)~,
\end{eqnarray}
where we employed a notation analogous to Eq.~\eqref{Eq:fd_mixed}, only for supercells with a single atom displaced.

We note that, to evaluate the integral in the last line of Eq.~\eqref{Eq:2ndeph_simple}, one could also replace the Kohn-Sham potential by the complete Kohn-Sham Hamiltonian $\hat{H}_{\rm KS}$. This is formally legitimate since the kinetic term does not depend on the atomic coordinates, and could be used to turn Eqs.~\eqref{Eq:fd_mixed} and \eqref{Eq:fd_diag} into expressions containing the Kohn-Sham eigenvalues of the supercell; this strategy was employed in Ref.~\citenum{Poliukhin_Marzari_2025} to evaluate linear electron-phonon coupling matrix elements. 

For validation purposes, we will also compute the linear electron-phonon matrix elements using the same finite-differences strategy as in Eqs.~\eqref{Eq:2ndeph_simple}-\eqref{Eq:fd_diag}. In the case of first-order couplings, the counterpart of these expressions is~\cite{Chaput_Tanaka_2019}:
\begin{eqnarray}\label{Eq:fd_linear}
    && g^{(1)}_{mn\nu}(\bk,\bq)
    = N_p\sum_{\k \a}
    \sqrt{ \frac{\hbar} {2M_\k \w_{\bq\nu}}}
    e_{\k \a,\nu}(\bq) \frac{1}{2 \d}
    \nonumber \\
    &&\hspace{20pt}\times [\hspace{5pt}\< \psi_{m\bk+\bq}| V_{\rm KS} (\tau_{\k\a 0}+\d) |\psi_{n\bk} \>_{\rm sc}\nonumber\\
    &&\hspace{28pt}- \< \psi_{m\bk+\bq}| V_{\rm KS} (\tau_{\k\a 0}-\d) |\psi_{n\bk} \>_{\rm sc}] + \mathcal{O}(\d^2).
\end{eqnarray}
By comparing Eq.~\eqref{Eq:fd_linear} and Eq.~\eqref{Eq:fd_diag}, we see that the evaluation of the second-order matrix elements provides all the ingredients needed to also compute the first-order matrix elements, at no extra cost.

The remaining question is how to efficiently evaluate integrals of the type:
  \begin{equation}\label{Eq:key_integral}
      \< \psi_{m\bk+\bq}| \tilde V_{\rm KS} |\psi_{n\bk} \>_{\rm sc},
  \end{equation}
where $\tilde V_{\rm KS}$ represents any one of the perturbed potentials appearing in Eqs.~\eqref{Eq:fd_mixed} and \eqref{Eq:fd_diag}, i.e., it corresponds to a supercell where one or two atoms have been slightly displaced from their equilibrium sites. To this end, our strategy is to combine:
\begin{itemize}
\item[(a)] the perturbed Kohn-Sham potential in the BvK supercell, and\\[-15pt]
\item[(b)] the unperturbed wavefunctions in the unit cell.
\end{itemize}
This strategy is advantageous for two reasons: (i) we completely avoid calculations of unoccupied Kohn-Sham states in the supercell; this choice affords us a considerable computational saving. (ii) Our unit-cell wavefunctions are directly amenable to Wannier interpolation (see Sec.~\ref{Sec:Wannier}), making the present approach fully compatible with the existing \textsc{EPW} infrastructure for electron-phonon calculations~\cite{Lee_Giustino_2023} and related codes that rely on Wannier-Fourier interpolation. 

\begin{figure}
    \centering
    \includegraphics[width=\linewidth]{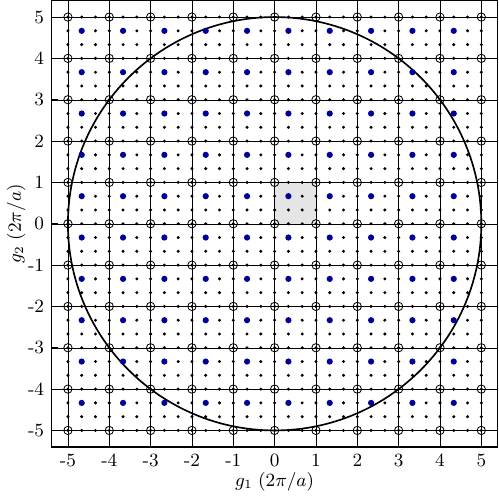}
    \caption{Relation between the reciprocal space of the unit cell and that of the BvK supercell.
    This example illustrates a two-dimensional square lattice with lattice parameter $a$ and BvK supercell of size 3$\times$3 unit cells. The black dots are the reciprocal lattice vectors
    of the BvK supercell, $\bG$. The circles are the reciprocal lattice vectors of the unit cell, $\bg$.
    The filled gray square is the unit-cell Brillouin zone, and the blue dot is a representative $\bk$ point inside the Brillouin zone. The other blue dots are the set of points $\bk+\bg$. For each such point, there is a unique $\bG$ in the supercell reciprocal lattice. The circle denotes the wavevector cutoff (in this example, $5 \times 2\pi/a$).}
    \label{Fig:gvecs}
\end{figure}

In the case of pseudopotential planewaves codes, the evaluation of Eq.~\eqref{Eq:key_integral} proceeds as follows. First, we decide on a uniform Brillouin zone sampling, say $N_1 \times N_2 \times N_3$ points. By denoting the primitive vectors of the reciprocal lattice via $\bb_1$, $\bb_2$, and $\bb_3$, the $\bk$-point grid is given by:
  \begin{equation}\label{Eq:kpoint_grid}
      \bk = \frac{k_1}{N_1} \bb_1 + \frac{k_2}{N_2}\bb_2 + \frac{k_3}{N_3}\bb_3, \quad k_\a = 0,1,\cdots,N_\a-1~.
  \end{equation}
Then, we compute Kohn-Sham (pseudo-)wavefunctions in the unit cell on this grid. In a basis of planewaves, these wavefunctions are represented as:
  \begin{equation}\label{Eq:psik}
      \psi_{n\bk}(\br) = \frac{1}{\sqrt{N_1 N_2 N_3 \Omega_{\rm uc}}}\sum_\bg c_{n\bk}(\bg) \exp[i(\bk+\bg)\cdot \br]~,
  \end{equation}
where $\Omega_{\rm uc}$ is the volume of the unit cell, and the reciprocal lattice vectors of the primitive unit cell are given by:
  \begin{equation}\label{Eq:gvecs}
      \bg = g_1 \bb_1 + g_2\bb_2 + g_3\bb_3, \quad \text{with}\,\, g_\a \,\,\text{integer}~.
  \end{equation}
Normalization of the wavefunctions in the BvK supercell implies $\sum_\bg |c_{n\bk}(\bg)|^2=1$.

The third step is to map the wavefunctions $\psi_{n\bk}$ into the reciprocal space of the BvK supercell. To this end, the supercell is constructed to have primitive (supercell) lattice vectors $\bA_\a = N_\a \ba_\a$ and is sampled at the $\Gamma$ point, so as to keep the electronic Hilbert spaces in the unit cell and in the supercell equivalent. Consequently, the primitive vectors of the reciprocal lattice of the BvK supercell are given by:
  \begin{equation}\label{Eq:Bvecs}
      \bB_1 = 2\pi \frac{\bA_2\times \bA_3}{\bA_1 \cdot \bA_2 \times \bA_3} =
        \frac{1}{N_1} 2\pi \frac{\ba_2\times \ba_3}{\ba_1 \cdot \ba_2 \times \ba_3}
        = \frac{\bb_1}{N_1}~, 
  \end{equation}
and similarly for the other components. These vectors generate the reciprocal lattice of the BvK supercell:
  \begin{equation}\label{Eq:Gvecs}
      \bG = \sum_\a G_\a \bB_\a = \sum_\a \frac{G_\a}{N_\a} \bb_\a, \quad \text{with}\,\, G_\a \,\,\text{integer}~.
  \end{equation}
Now we remark that the set of vectors $\bk+\bg$ and the set of vectors $\bG$ span the same space. In fact, by combining Eqs.~\eqref{Eq:kpoint_grid}, \eqref{Eq:gvecs}, \eqref{Eq:Bvecs}, \eqref{Eq:Gvecs}, we see that $\bk + \bg$ and $\bG$ coincide whenever:
  \begin{equation}\label{Eq:corresp}
     G_1 = N_1 g_1 + k_1,\quad G_2 = N_2 g_2 + k_2, \quad G_3 = N_3 g_3 + k_3~.      
  \end{equation}
This equation allows us to map the planewaves coefficients $c_{n\bk}(\bg)$ of the unit-cell wavefunction into the planewaves coefficients of the corresponding wavefunction in the BvK supercell, see Fig.~\ref{Fig:gvecs}. Since the planewaves cutoff is imposed on $|\bk+\bg|$ in the unit cell, and on $|\bG|$ in the BvK supercell, there is an exact one-to-one correspondence between the planewave coefficients in the two representations, without any loss of information.

Using this correspondence, we rewrite Eq.~\eqref{Eq:key_integral} using the \textit{resolution of identity in reciprocal space}:
  \begin{eqnarray}\label{Eq:key_integral2}
     \hspace{-10pt} \< \psi_{m\bk+\bq}| \tilde V_{\rm KS} |\psi_{n\bk} \>_{\rm sc} &=&
       \sum_{\bG,\bG'} \< \psi_{m\bk+\bq}| \bG\>_{\rm sc} \nonumber \\ 
          &\times& 
    \<\bG |\tilde V_{\rm KS} |\bG'\>_{\rm sc}\<\bG'|\psi_{n\bk} \>_{\rm sc}~,\,\,    
  \end{eqnarray}
where $|\bG\>$ denotes the normalized planewave $(N_1 N_2 N_3\Omega_{\rm uc})^{-1/2}\exp(i\bG\cdot\br)$ when evaluated in the position representation. Upon replacing Eq.~\eqref{Eq:psik} in this expression, we obtain:
  \begin{eqnarray}\label{Eq:key_integral3}
     &&\< \psi_{m\bk+\bq}| \tilde V_{\rm KS} |\psi_{n\bk} \>_{\rm sc} =
       \sum_{\bG}   c^*_{m\bk+\bq}(\bG-\bk-\bq) \nonumber \\  
   &&\hspace{30pt}\times \sum_{\bG'}\<\bG |\tilde V_{\rm KS} |\bG'\>_{\rm sc} \,  c_{n\bk}(\bG'-\bk)~.    
  \end{eqnarray}
This step completes our procedure. In practice, we compute the set of wavefunctions $\psi_{n\bk}$ in the unit cell, and we load the Fourier coefficients in the BvK supercell calculations by mapping onto the reciprocal space of the supercell via the rule $\bg \rightarrow \bG=\bk+\bg$. From this point onward, the evaluation of the integrals proceeds as in standard planewaves codes. For example, in \textsc{Quantum ESPRESSO} one applies the Hartree potential in reciprocal space and the local (ionic plus exchange and correlation) potential in real space using fast Fourier transforms. Importantly, this formalism can be used with either semilocal functionals or with fully nonlocal functionals~\cite{Kronik_Baer_2012} without any restrictions.
\red{It should be noted that a similar strategy has been adopted in the calculation of electron-defect couplings~\cite{Lu_Bernardi_2022}, where the evaluations of supercell wavefunctions have been effectively avoided.}
Together, Eqs.~\eqref{Eq:2ndeph_simple}-\eqref{Eq:fd_diag} and \eqref{Eq:key_integral3} constitute the central result of the present work.

The generalization of Eqs.~\eqref{Eq:2ndeph_simple}-\eqref{Eq:fd_diag}, \eqref{Eq:key_integral3} to higher-order nonlinear electron-phonon matrix elements is formally straightforward. The approach still relies on the evaluation of Eq.~\eqref{Eq:key_integral3}, and the only thing that changes is the linear combination of perturbed potential needed to evaluate higher-order derivatives. As an example, the third-order electron-phonon matrix elements can be written in a form analogous to Eq.~\eqref{Eq:2ndeph_simple}:
\begin{eqnarray}
    &&g^{(3)}_{mn\nu\nu'\nu''}(\bk,\bq,\bq',\bq'') = 
    \frac{N_p}{3!}  \nonumber \\ && \times
    \sum_{\k \a, \k' \a',\k'\a''}
    \sqrt{ \frac{\hbar} {2M_\k \w_{\bq\nu}}}\sqrt{ \frac{\hbar} {2M_{\k'} \w_{\bq'\nu'}}}
    \sqrt{ \frac{\hbar} {2M_{\k''} \w_{\bq''\nu''}}}
    \nonumber \\ && \times\, 
    e_{\k \a,\nu}(\bq)e_{\k'\a',\nu'}(\bq')e_{\k''\a'',\nu''}(\bq'') \sum_{pp'} e^{i(\bq\cdot\bR_p+\bq'\cdot\bR_{p'})}
    \nonumber \\
    &&\times   \< \psi_{m\bk+\bq+\bq'+\bq''}| \frac{\partial^3 V_{\rm KS}}{\partial \tau_{\k\a p} \partial \tau_{\k'\a' p'} \partial \tau_{\k''\a'' 0}} |\psi_{n\bk}\>_{\rm sc}~.
    \label{Eq:3rdeph_simple}    
\end{eqnarray}
The evaluation of the integrals in the last line proceeds as in Eq.~\eqref{Eq:key_integral3}, only using third-order finite-difference formulas. The remaining challenge with this approach is that the number of BvK supercell calculations scales with $(3N_{\rm at})^3 N_p^2$, which is prohibitively expensive even for the smallest supercells. However, this challenge can be overcome by accepting some level of approximation; this is possible when using the RIA or the finite-range approximations to be discussed in Sec.~\ref{Sec:RIA}.

\subsection{Rigid-ion approximation for the second-order matrix element}
\label{Sec:RIA}

The RIA has been used in prior work to perform approximate calculations of second-order electron-phonon coupling matrix elements~\cite{Allen_Heine_1976, Allen_Cardona_1981, Giustino_Cohen_2010, Ponce_Gonze_2015, Lihm_Park_2020, Ponce_Park_2025}. 
The RIA is in essence an extreme short-range approximation to Eq.~\eqref{Eq:2ndeph_simple}, whereby one restrict the sum over lattice vectors to $\bR_p=0$, and neglects mixed second derivatives in the atomic coordinates ($\k\ne\k'$):
\begin{eqnarray}
    &&g^{(2),\text{RIA}}_{mn\nu\nu'}(\bk,\bq,\bq') = 
    \frac{1}{2} N_p\!\!\!
    \sum_{\k \a, \a'}
    \sqrt{ \frac{\hbar} {2M_\k \w_{\bq\nu}}}\sqrt{ \frac{\hbar} {2M_{\k} \w_{\bq'\nu'}}}
    \nonumber \\ && \hspace{10pt}\times\, 
    e_{\k \a,\nu}(\bq)e_{\k\a',\nu'}(\bq')
    \nonumber \\
    &&\hspace{10pt}\times  \< \psi_{m\bk+\bq+\bq'}| \frac{\partial^2 V_{\rm KS}}{\partial \tau_{\k\a 0} \partial \tau_{\k\a' 0}} |\psi_{n\bk}\>_{\rm sc}~.
    \label{Eq:2ndeph_RIA}    
\end{eqnarray}
This approximation is referred to as ``rigid ion'' because it becomes exact if the potential can be expressed a sum of independent ionic pseudopotentials $V_\k(\br)$ that move rigidly with the ions. For example, in the case of the ionic pseudopotential used in DFT calculations, one has:
  \begin{equation}\label{Eq:RI}
      V_{\rm ion} = \sum_{\k p} V_\k(\br-\btau_\k-\bR_p)~,
  \end{equation}
where we have considered a local potential for notational simplicity. 
In this case, the second derivatives of the potential read:
  \begin{equation}\label{Eq:RI2}
      \frac{\partial^2 V_{\rm ion}(\br)}{\partial \tau_{\k\a p} \partial \tau_{\k'\a' p'}} =
      \d_{\k\k'}\d_{pp'}\frac{\partial^2 V_\k(\br-\btau_\k-\bR_p)}{\partial r_\a \partial r_{\a'}}~,
  \end{equation}
therefore they vanish identically when $p\ne p'$ or $\k\ne\k'$. In fact, it is not a coincidence that the historical origin of the RIA is rooted in the empirical pseudopotential method~\cite{Cohen_Bergstresser_1966, Allen_Cohen_1969}; in that method, the potential is precisely of the form of Eq.~\eqref{Eq:RI}, and there is no self-consistent contribution from the electron density as in standard DFT.

Conceptually, the RIA can be thought of as a statement about the inverse dielectric matrix. In fact, the variation of the Kohn-Sham potential can formally be expressed as~\cite{Giustino_2017}:
\begin{equation}\label{Eq:RI3}
  \frac{\partial V_{\rm KS}(\br)}{\partial \tau_{\k\a 0}} =
  \int \!d\br' \epsilon^{-1}(\br,\br') \frac{\partial V_{\rm ion}(\br')}{\partial \tau_{\k\a 0}}~,    
\end{equation}
where $\epsilon^{-1}$ denotes the inverse electronic dielectric matrix in DFT. Since this dielectric matrix is a parametric function of the atomic coordinates, using Eqs.~\eqref{Eq:RI}-\eqref{Eq:RI3} we can write the second derivative of the Kohn-Sham potential as:
\begin{eqnarray}\label{Eq:RIA-decomp}
  &&\hspace{-20pt}\frac{\partial^2 V_{\rm KS}(\br)}{\partial \tau_{\k\a 0} \partial \tau_{\k'\a' p'}}
  = -\int \!d\br' \frac{\partial \epsilon^{-1}(\br,\br')}{\partial \tau_{\k'\a' p'}} \frac{\partial V_\k(\br'-\btau_\k)}{\partial r_\a} \nonumber \\
&&\hspace{5pt}+ \d_{\k\k'}\d_{pp'}\int \!d\br' \epsilon^{-1}(\br,\br') 
  \frac{\partial^2 V_\k(\br-\btau_\k-\bR_p)}{\partial r_\a \partial r_{\a'}}.\hspace{5pt}
\end{eqnarray}
The second line of this equation does not contain mixed derivatives, hence it behaves like a rigid-ion contribution, precisely as in Eq.~\eqref{Eq:RI2}. The source of non-rigid-ion contributions is therefore the term in the first line containing the variations $\partial \epsilon^{-1}(\br,\br')/\partial \tau_{\k'\a' p'}$. Since these variations are significant only in the neighborhood of $\btau_{\k'p'}$ and vanish away from it~\cite{Giustino_Pasquarello_2005}, this term contributes to the integral only when $\btau_{\k'p'}$ is close to $\btau_{\k0}$. In Sec.~\ref{Sec:RIA_better} we exploit this observation to develop more accurate methods beyond the RIA.

In practice, the RIA is useful to  drastically reduce the computational cost if one is willing to accept a certain level of approximation. In fact, the evaluation of Eq.~\eqref{Eq:2ndeph_RIA} requires $3^2 N_{\rm at}$ supercell calculations, to be contrasted with the much higher number $(3N_{\rm at})^2 N_p$ of evaluations required by the full expression in Eq.~\eqref{Eq:2ndeph_simple}. Thus, the computational saving afforded by the RIA is a factor $N_{\rm at}N_p$; for example, in the case of a 3$\times$3$\times$3 BvK supercell with 2 atoms per primitive cell, the RIA yields a 54-fold reduction in computational cost.

The same reasoning applies to higher-order electron-phonon coupling matrix elements, such as the third-order matrix element in Eq.~\eqref{Eq:3rdeph_simple}. In that case, the computational cost is reduced from $(3N_{\rm at})^3 N_p^2$ supercell calculations to only
$3^3 N_{\rm at}$ calculations, which corresponds to a saving with a factor of $(N_{\rm at}N_p)^2$; using the previous example of a 3$\times$3$\times$3 BvK supercell with 2 atoms per cell, this factor corresponds to a nearly 3000-fold reduction in computational cost. Given the enormous saving afforded by this approximation even for the simplest systems, it is worth investigating the error incurred by the RIA, as well as whether we can devise intermediate approximations that reduce the number of required derivatives in Eq.~\eqref{Eq:2ndeph_simple} without losing much accuracy. We address these questions in Sec.~\ref{Sec:RIA_bad}.

Historically, the RIA has been instrumental to obtaining the Debye-Waller energy shift shown in Eq.~\eqref{Eq:dw_selfe}. In fact, prior to the present work, direct calculations of all the second-order matrix elements required in Eq.~\eqref{Eq:dw_selfe} had not been possible; to circumvent this limitation, it is standard practice to replace the matrix elements $g^{(2)}_{nn\nu\nu}(\bk,\bq,-\bq)$ by their RIA approximations~\cite{Allen_Heine_1976,Giustino_Cohen_2010,Ponce_Park_2025}:
\begin{equation}
    \label{Eq:dw_selfe_ria}
    \Delta \varepsilon_{n\bk}^{\rm DW, RIA} 
    = \frac{1}{N_p}\sum_{\bq\nu} g^{(2),\text{RIA}}_{nn\nu\nu}(\bk,\bq,-\bq)~,
\end{equation}
where $g^{(2),\text{RIA}}_{nn\nu\nu}(\bk,\bq,-\bq)$ is given by Eq.~\eqref{Eq:2ndeph_RIA}.
The advantage of this replacement is that, as a consequence of the acoustic sum rule, the sum $\sum_{\bq\nu}$ over the second-order RIA matrix elements can be recast into an equivalent sum of first-order, exact matrix elements $g^{(1)}_{mn\nu}(\bk,\bq)$~\cite{Allen_Heine_1976,Giustino_Cohen_2010}. This replacement effectively allows one to completely bypass the explicit evaluation of second-order matrix elements. In related approaches, it has been shown that the sum in Eq.~\eqref{Eq:dw_selfe_ria} can be recast in terms of matrix elements of commutators involving the first variation of the Kohn-Sham potential and the momentum operator~\cite{Lihm_Park_2020}; and that the summation can in turn be replced by a self-consistent Sternheimer equation~\cite{Gonze_Cote_2011, Ponce_Gonze_2015}.
The common denominator to these approaches is that they evaluate \textit{sums over a subset} of all possible matrix elements. They do not provide access to individual matrix elements $g^{(2),\text{RIA}}_{mn\nu\nu'}(\bk,\bq,\bq')$ or $g^{(2)}_{mn\nu\nu'}(\bk,\bq,\bq')$.

\subsection{Finite-range approximation for the second-order matrix element}
\label{Sec:RIA_better}

Equation~\eqref{Eq:RIA-decomp} suggests that one might be able to improve upon the RIA by systematically including contributions from atoms $(\k'p')$ that are further away from $(\k 0)$. It is therefore natural to consider a ``finite-range approximation'' (FRA) by introducing a spatial cutoff $d$ in Eq.~\eqref{Eq:2ndeph_simple}, so that only mixed derivatives involving atoms $(\k p)$ and $(\k'0)$ that are within the distance $d$ are retained:
\begin{eqnarray}
    &&g^{(2),\text{FRA}}_{mn\nu\nu'}(\bk,\bq,\bq') = 
    \sum_{\k \k'p} \theta(d-|\bR_p + \btau_{\k0}-\btau_{\k'0}|)
    \nonumber \\ 
    &&\hspace{10pt}\times \frac{N_p}{2} \sum_{\a \a'}
    \sqrt{ \frac{\hbar} {2M_\k \w_{\bq\nu}}}\sqrt{ \frac{\hbar} {2M_{\k'} \w_{\bq'\nu'}}}
    \nonumber \\ && \hspace{10pt}\times\, 
    e_{\k \a,\nu}(\bq)e_{\k'\a',\nu'}(\bq')
    \nonumber \\
    &&\hspace{10pt}\times   e^{i\bq\cdot\bR_p} \< \psi_{m\bk+\bq+\bq'}| \frac{\partial^2 V_{\rm KS}}{\partial \tau_{\k\a p} \partial \tau_{\k'\a' 0}} |\psi_{n\bk}\>_{\rm sc}~.
    \label{Eq:2ndeph_simple_FRA}    
\end{eqnarray}
Here, the Heaviside theta function restricts the sums to those atoms that satisfy the condition
$|\bR_p+\btau_{\k0}-\btau_{\k'0}|\le d$. In practice, since the the most time-consuming part of the calculation is the evaluation of the integrals in the last line, we evaluate this expression by first constructing a lookup table of the allowed combinations of $(\k,\k',p)$, and then we calculate the integrals only for these combinations.
Even though this search for neighbors is straightforward, to better illustrate the concept we briefly discuss the cases of diamond and LiF, which we use as test cases in Sec.~\ref{Sec:LongerRange}.

The primitive vectors of the diamond lattice are
$\ba_1 = a (\bu_y + \bu_z)/2$,   
$\ba_2 = a (\bu_x + \bu_z)/2$, and
$\ba_3 = a (\bu_x + \bu_y)/2$, where $a$ is the lattice parameter are $\bu_x$, $\bu_y$, and $\bu_z$ are Cartesian unit vectors. 
The positions of the two carbon atoms in the reference cell are
$\btau_1 =0$ (labeled as atom 1) and $\btau_2 = a(\bu_x + \bu_y +\bu_z)/4$ (labeled as atom 2). Therefore, using $\bR_p = n_1\ba_1 + n_2\ba_2+n_3\ba_3$, the distance $d_{\k\k'p}$ between atoms $(\k p)$ and $(\k'0)$ is given by: 
\begin{eqnarray}
  \left(\!\frac{d_{\k\k'p}}{a}\!\right)^{\!\!2} 
    &\!\!=\!\!&
    \left[(n_1-n_2)^2+(n_2-n_3)^2+(n_3-n_1)^2\right]/12 \nonumber \\
    &\!\!+\!\!&\left[n_1+n_2+n_3+3(\k-\k')/4\right]^2/3~,
\end{eqnarray}
By cycling through $n_1$, $n_2$, and $n_3$, one obtains the following nearest-neighbor shells:
\begin{itemize}
    \item[(a)] 0th-NN shell: This subset corresponds to setting $d=0$, and only includes the combinations $\k=\k'$ and $(n_1,n_2,n_3)=(0,0,0)$; therefore, only 
    $2\times3\times3=18$
    integrals are retained in Eq.~\eqref{Eq:2ndeph_simple_FRA}. 
    This choice corresponds to the RIA.
    \item[(b)] 1st-NN shell: This subset corresponds to setting $d=\sqrt{3}a/4$, and for each carbon atom it includes its four nearest neighbors within the same cell or in the surrounding cells: 
    $(n_1,n_2,n_3)=(0,0,0)$, $(0,0, 1)$, $(0, 1,0)$, $( 1,0,0)$ for $(\k,\k')=(1,2)$; and
    $(n_1,n_2,n_3)=(0,0,0)$, $(0,0,-1)$, $(0,-1,0)$, $(-1,0,0)$ for $(\k,\k')=(2,1)$. We have a total of $2\times4\times3\times3=72$ integrals in Eq.~\eqref{Eq:2ndeph_simple_FRA}.
    \item[(c)] 2nd-NN shell: This subset corresponds to setting $d=\sqrt{2}a/2$. For each carbon atom it includes its twelve second-nearest neighbors: $\k=\k'$ and 
    $(n_1,n_2,n_3)=(\pm 1,0,0)$ and cyclic permutations, or $\k=\k'$ and
    $(n_1,n_2,n_3)=(\pm 1,\mp 1,0)$ and cyclic permutations, for a total of $2\times12\times3\times3=216$ integrals in Eq.~\eqref{Eq:2ndeph_simple_FRA}.
\end{itemize}
A similar analysis can be carried out for LiF. In this case, the primitive lattice vectors are the same as in diamond, while the basis is given by
$\btau_1 =0$ (labeled as atom 1) and $\btau_2 = a(\bu_x + \bu_y + \bu_z)/2$ (labeled as atom 2). 
In this case we have the following shells of neighbors:
\begin{itemize}
    \item[(a)] 0th-NN shell: Also in this case, this is the RIA and amounts to evaluating $2\times3\times3=18$ integrals corresponding $\k=\k'$ and $(n_1,n_2,n_3)=(0,0,0)$ in Eq.~\eqref{Eq:2ndeph_simple_FRA}.
    \item[(b)] 1st-NN shell: This subset corresponds to setting $d=a/2$, and for each Li (F) atom it includes its six F (Li) nearest neighbors in the surrounding cells: 
    $(n_1,n_2,n_3)=(0,0,1)$, $(0,1,0)$, $(1,0,0)$, $(0,1,1)$, $(1,1,0)$, $(1,0,1)$ for $(\k,\k')=(1,2)$;
    and 
    $(n_1,n_2,n_3)=(0,0,-1)$, $(0,-1,0)$, $(-1,0,0)$, $(0,-1,-1)$, $(-1,-1,0)$, $(-1,0,-1)$ for
    $(\k,\k')=(2,1)$.
    In this case we have to evaluate a total of $2\times6\times3\times3=108$ integrals in Eq.~\eqref{Eq:2ndeph_simple_FRA}.
    \item[(c)] 2nd-NN shell: This subset corresponds to setting $d=\sqrt{2}a/2$. For each Li or F atom, it includes its twelve second-nearest neighbors: $\k=\k'$ and 
    $(n_1,n_2,n_3)=(\pm 1,0,0)$ and cyclic permutations, or 
    $(n_1,n_2,n_3)=(\pm 1,\mp 1,0)$ and cyclic permutations, for a total of $2\times12\times3\times3=216$ integrals in Eq.~\eqref{Eq:2ndeph_simple_FRA}. This shell structure is the same as for diamond.
\end{itemize}
In Sec.~\ref{Sec:LongerRange} we test the accuracy of these choices and compare to the RIA.

\subsection{\textit{Ab initio} polaron equations with second-order electron-phonon couplings}
\label{Sec:Polaroneq}

In this section we generalize the \textit{ab initio} polaron equations~\cite{Sio_Giustino_2019a, Sio_Giustino_2019b,Dai_Giustino_2025a} to the case of second-order electron-phonon couplings. In their original formulation, the polaron equations involve the first-order couplings $g^{(1)}_{mn\nu}(\bk,\bq)$; we refer to these equations as the ``linear polaron equations''. Here, we extend these equations to include the effect of second-order matrix elements $g^{(2)}_{mn\nu\nu'}(\bk,\bq,\bq')$. We refer to this generalization as the ``quadratic polaron equations''.

Following Ref.~\cite{Dai_Giustino_2026}, the DFT total energy functional of a system with one excess electron can be written compactly as:
\begin{align}
    \label{Eq:Etot_Sadigh}
    E^{N+1}[\psi_{\rm p}(\br), \btau] = 
    E^{N}(\btau) + \bra{\psi_{\rm p}} \hH_{\rm KS}(\btau) \ket{\psi_{\rm p}}~.
\end{align}
In this expression, $E^N$ and $E^{N+1}$ indicate the total energies of the neutral system with $N$ electron and the charged system with $N+1$ electrons, respectively; $\btau$ is a collective coordinate that specifies the position of all the atoms in the system; $\psi_{\rm p}$ is the wavefunction of the excess electron, which is identified with the polaron state; and $\hH_{\rm KS}$ is the Kohn-Sham Hamiltonian of the $N$-electron system. In Eq.~\eqref{Eq:Etot_Sadigh}, the polaron self-interaction error has already been removed; a detailed discussion of the approximations underlying this equation can be found in Ref.~\citenum{Dai_Giustino_2026}.

Let us denote the equilibrium configuration of the $N$-electron system by $\btau_0$. Upon introducing the excess electron, the atoms are displaced by $\Delta \btau$, so that $\btau = \btau_0+\Delta \btau$. The linear polaron equations are obtained by performing a Taylor expansion of Eq.~\eqref{Eq:Etot_Sadigh} to the lowest order in $\Delta \btau$ that allows electron localization. This choice corresponds to retaining second-order terms for $E^{N}(\btau)$ and first-order terms for $\bra{\psi_{\rm p}} \hH_{\rm KS}(\btau) \ket{\psi_{\rm p}}$:
\begin{align}
    &E^N(\btau) \approx E^N(\btau_0) 
    + \frac{1}{2}\sum_{\substack{\k \a p\\\k'\a' p}} C_{\k p \a, \k' p' \a'} \Delta \tau_{\k \a p} \Delta \tau_{\k'\a'p'}, 
    \label{Eq:harmonic}
    \\
    &\hH_{\rm KS}(\btau) \approx \hH_{\rm KS}(\btau_0) + \sum_{\k \a p}\frac{\D V_{\rm KS}}{\D \tau_{\k \a p}} \Delta \tau_{\k \a p}~. \label{Eq:linear_H}
\end{align}
Upon replacing these expressions inside Eq.~\eqref{Eq:Etot_Sadigh} and performing variational minimization of the energy with respect to $\psi_{\rm p}$ and $\Delta\btau$ subject to the normalization constraint for the wavefunction, one obtains~\cite{Sio_Giustino_2019a}:
\begin{align}
     &\bigg[\hH_{\rm KS}(\btau_0) + \sum_{\k \a p}\frac{\D V_{\rm KS}}{\D \tau_{\k \a p}} \Delta \tau_{\k \a p}
     \bigg] \psi_{\rm p}(\br) = \ve\, \psi_{\rm p}(\br), \label{Eq:linear_plrneq_1}
     \\
     &\Delta \tau_{\k\a p} = 
    -\hspace{-6pt}\sum_{\k' \a' p'} C_{\k \a p, \k' \a' p'} 
    \!\int\!\! d\br \frac{\D V_{\rm KS}}{\D \tau_{\k' \a' p'}}  |\psi_{\rm p}(\br)|^2.
    \label{Eq:linear_plrneq_2}
\end{align}
In Eq.~\eqref{Eq:linear_plrneq_1}, $\ve$ is the Lagrange multiplier corresponding to the normalization constraint, and carries the physical meaning of vertical excitation energies of the polaron~\cite{Dai_Giustino_2025a}. 

Equations~\eqref{Eq:linear_plrneq_1}-\eqref{Eq:linear_plrneq_2} have been used in a number of studies to investigate small and large polarons in materials~\cite{Sio_Giustino_2019a,Sio_Giustino_2019b,Vasilchenko_Gonze_2022,Vasilchenko_Gonze_2024,Vasilchenko_Gonze_2025,Dai_Giustino_2024a,Dai_Giustino_2026,Dai_Giustino_2024b,Sio_Giustino_2023,Lafuente_Giustino_2024,Luo_Giustino_2026}; however, they clearly rely on an unbalanced Taylor expansion of the total energy (second order) and the Hamiltonian (first order). 
A formally more rigorous way to proceed is to carry out both expansions to the second order.
To this end, we replace Eq.~\eqref{Eq:linear_H} by the second-order expansion:
\begin{eqnarray}
    \hspace{-20pt}\hH_{\rm KS}(\btau) 
    &\approx& \hH_{\rm KS}(\btau_0) + \sum_{\k \a p}\frac{\D V_{\rm KS}}{\D \tau_{\k \a p}} \Delta \tau_{\k \a p}
    \nonumber \\
    &+&\frac{1}{2}\sum_{\substack{\k \a p \\ \k' \a' p'}} \frac{\D^2 V_{\rm KS}}{\D \tau_{\k \a p}\D \tau_{\k' \a' p'}}
    \dtau_{\k \a p} \dtau_{\k' \a' p'}
    \label{Eq:second_H}~.
\end{eqnarray}
The last line of this expression introduces quadratic electron-phonon couplings in the picture. If we replace Eq.~\eqref{Eq:second_H} and \eqref{Eq:harmonic} inside Eq.~\eqref{Eq:Etot_Sadigh} and repeat the variational minimization, we find the \textit{quadratic polaron equations}:
\begin{align}
    &\bigg[
    \hH_{\rm KS}(\btau_0) 
    + \sum_{\k \a p}\frac{\D V_{\rm KS}}{\D \tau_{\k \a p}} \Delta \tau_{\k \a p}
    \nonumber \\
    &+\frac{1}{2}\sum_{\substack{\k \a p \\ \k' \a' p'}} \frac{\D^2 V_{\rm KS}}{\D \tau_{\k \a p}\D \tau_{\k' \a' p'}}
    \dtau_{\k \a p} \dtau_{\k' \a' p'}
    \bigg] \psi_{\rm p}(\br) = \ve\, \psi_{\rm p}(\br)
    \label{Eq:2nd_plrneq_1},  
    \\
    &
    \sum_{\substack{\k' \a' p'}} 
    C_{\k \a p, \k' \a' p'} 
    \dtau_{\k' \a' p'}
    +
    \int_{\rm sc}d\br
    \frac{\D V_{\rm KS}}{\D \tau_{\k \a p}}\abs{\psi_{\rm p}(\br)}^2
    \nonumber \\
    &+\sum_{\k' \a' p'}
    \int_{\rm sc}d\br \frac{\D^2 V_{\rm KS}}{\D \tau_{\k \a p}\D \tau_{\k' \a' p'}} \Delta \tau_{\k' \a' p'}
    \abs{\psi_{\rm p}(\br)}^2
    =0.
    \label{Eq:2nd_plrneq_2}  
\end{align}
As for the case of the linear polaron equations, it is advantageous to express the wavefunction $\psi_{\rm p}(\br)$ and the atomic displacements $\dtau_{\k \a p}$ in terms of the Kohn-Sham states $\psi_{n\bk}$ and vibrational eigenmodes $e_{\k \a,\nu}(\bq)$ of the unperturbed crystal~\cite{Sio_Giustino_2019b}:
\begin{align}
    &\psi_{\rm p}(\br) = \frac{1}{\sqrt{N_p}} \sum_{n\bk}
    A_{n\bk} \psi_{n\bk}(\br),
    \label{Eq:Psi_decom}
    \\
    &
    \dtau_{\k \a p}
    =
    -\frac{2}{N_p} \sum_{\bq\nu} B^*_{\bq\nu}
    \sqrt{ \frac{\hbar} {2M_\k \w_{\bq\nu}}}
    e_{\k \a,\nu}(\bq)
    e^{i\bq\cdot\bR_p}~,
    \label{Eq:dtau_decom}
\end{align}
where the expansion coefficients $A_{n\bk}$ and $B_{\bq\nu}$ are to be determined. Upon replacing Eqs.~\eqref{Eq:Psi_decom}-\eqref{Eq:dtau_decom} inside Eqs.~\eqref{Eq:2nd_plrneq_1} and \eqref{Eq:2nd_plrneq_2}, we find:
\begin{align}
    \label{Eq:2nd_plrneq_1_rec}
    &(\ve -\ve_{n\bk}) A_{n\bk}+\frac{2}{N_p}\sum_{m}^{M_1}  
    \sum_{\bq\nu} A_{m\bk+\bq} 
    g^{(1),*}_{mn\nu}(\bk,\bq)B_{\bq\nu}
    \nonumber \\
    &\hspace{10pt}=\frac{4}{N_p^2} \sum_{m}^{M_2}\sum_{\bq\nu,\bq'\nu'}
    A_{m\bk+\bq+\bq'} 
    g^{(2),*}_{mn\nu\nu'}(\bk,\bq,\bq')B_{\bq\nu} B_{\bq'\nu'},\\
    \label{Eq:2nd_plrneq_2_rec}
    &\frac{1}{N_p} \sum_{mn}^{M_1}\sum_{\bk} A^*_{m\bk+\bq} g^{(1),*}_{mn\nu}(\bk,\bq)A_{n\bk} - \hbar \w_{\bq\nu} B_{\bq\nu}\nonumber \\
    &=\frac{4}{N_p^2} \sum_{mn}^{M_2} \sum_{\bk,\bq'\nu'} A_{m\bk-\bq+\bq'} 
    g^{(2),*}_{mn\nu\nu'}(\bk,-\bq,\bq')A^*_{n\bk} B_{\bq'\nu'} .
\end{align} 
These expressions constitute the quadratic polaron equations in reciprocal space. If we set to zero the second-order couplings $g^{(2)}_{mn\nu\nu'}(\bk,\bq,\bq')$ in both equations, we recover the linear polaron equations of Refs.~\citenum{Sio_Giustino_2019b}.

In Eqs.~\eqref{Eq:2nd_plrneq_1_rec} and \eqref{Eq:2nd_plrneq_2_rec} we explicitly indicate the ranges $M_1$ and $M_2$ for the band sums involving first-order and second-order matrix elements, respectively. In principle, one should use $M_1=M_2 \rightarrow \infty$; in practice, since calculations of the second-order couplings are computationally demanding, it is advantageous to use $M_2<M_1$. The convergence of polaron calculations with respect to these parameters is investigated in App.~\ref{App:band_conv}.

At the end of this section, it is worth emphasizing that the solution of the quadratic polaron equations, Eqs.~\eqref{Eq:2nd_plrneq_1_rec} and \eqref{Eq:2nd_plrneq_2_rec}, requires knowledge of the entire set of second-order couplings $g^{(2)}_{mn\nu\nu'}(\bk,\bq,\bq')$. This situation is qualitatively different and significantly more challenging than the study of the Debye-Waller energy shifts in Eq.~\eqref{Eq:dw_selfe}, where only diagonal matrix elements (with $m=n$, $\nu'=\nu$, $\bq'=-\bq$) are needed.

\subsection{Wannier interpolation of the second-order electron-phonon coupling matrix elements}
\label{Sec:Wannier}

\begin{figure}
    \centering    \includegraphics[width=0.99\linewidth]{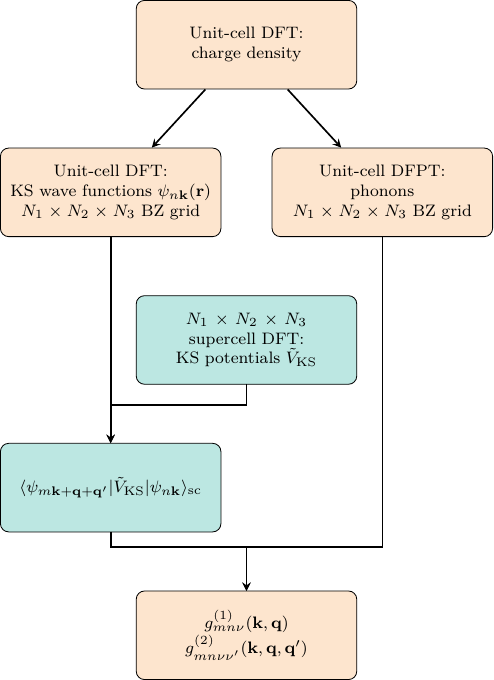}
    \caption{
    Workflow for calculating second-order electron-phonon coupling matrix elements via finite differences. The orange boxes indicate unit-cell DFT calculations, and the blue boxes indicate supercell DFT calculations. The Brillouin zone (BZ) grid for the unit-cell calculations equals the number of unit cells in the BvK supercell.}
    \label{Fig:Workflow}
\end{figure}

It is well known that many calculations of properties derived from electron-phonon couplings, from transport to superconductivity, require a fine sampling of the Brillouin zone in order to achieve numerical convergence~\cite{Giustino_2017}. Since each wavevector $\bq$ in $g^{(1)}_{mn\nu}(\bk,\bq)$ necessitates a separate DFPT or supercell calculation, sampling the Brillouin zone via direct calculations is not possible in practice. To circumvent this bottleneck, in the case of linear electron-phonon couplings efficient interpolation strategies have been developed, including Wannier-Fourier interpolation~\cite{Giustino_Louie_2007}, interpolation of the local potential response~\cite{Eiguren_Ambrosch-Draxl_2008}, and interpolation using atomic orbitals~\cite{Agapito_Buongiorno_2013, Gunst_Brandbyge_2016}. Since nonlinear electron-phonon couplings are even more demanding computationally than their linear counterparts, we can expect that the use of electron-phonon interpolation techniques will become essential in calculations of nonlinear electron-phonon interactions. In anticipation of these developments, we here derive the equations needed to perform Wannier-Fourier interpolation of second-order electron-phonon couplings. A similar strategy can be used for higher-order couplings.

Wannier functions are defined in terms of the Bloch states $\psi_{n\bk}(\br)$ via~\cite{Marzari_Vanderbilt_1997}: 
\begin{align}
    \label{Eq:b2w_wfc}
    w_{mp}(\br) = 
    \frac{1}{N_p} \sum_{n\bk}
    e^{-i\bk \cdot \bR_p} U_{nm\bk}
    \psi_{n\bk}(\br)~,
\end{align}
where $w_m(\br)$ is a Wannier function located in the reference unit cell and normalized in the BvK supercell; $U_{nm\bk}$ is a unitary matrix that ensures a smooth connection between Bloch states pertaining to neighboring $\bk$-points. The matrix $U_{mn\bk}$ is determined by requiring that the spatial spread of the Wannier functions be as small as possible, leading to the notion of maximally-localized Wannier functions (MLWFs)~\cite{Marzari_Vanderbilt_1997,Souza_Vanderbilt_2001}. The inverse relation of Eq.~\eqref{Eq:b2w_wfc} is: 
\begin{align}
    \label{Eq:w2b_wfc}
    \psi_{n\bk}(\br) = 
    \sum_{mp}
    e^{i\bk \cdot \bR_p} U_{mn\bk}^\dagger
    w_{mp}(\br)~.
\end{align}
By using this last relation inside Eq.~\eqref{Eq:2ndeph_simple}, we obtain: 
\begin{eqnarray}\label{Eq:w2b_eph}
   &&g^{(2)}_{mn\nu\nu'}(\bk,\bq,\bq') = 
    \frac{N_p}{2} \!\!\!
    \sum_{\k \a, \k' \a'}
    \sqrt{ \frac{\hbar} {2M_\k \w_{\bq\nu}}}\sqrt{ \frac{\hbar} {2M_{\k'} \w_{\bq'\nu'}}}
    \nonumber \\ && \hspace{10pt}\times\, 
    e_{\k \a,\nu}(\bq)e_{\k'\a',\nu'}(\bq') \sum_{m'n'} 
     U_{mm'\bk+\bq+\bq'}U_{n'n\bk}^\dagger  \nonumber \\ 
     &&\hspace{10pt} \times \sum_{pqr}  
     e^{i[\bk \cdot \bR_r + \bq\cdot\bR_p -(\bk+\bq+\bq') \cdot \bR_{q}]} 
    \nonumber \\
    &&\hspace{10pt}\times  g^{(2)}_{m'n',\k\a,\k'\a'}(\bR_p,\bR_q,\bR_r),
\end{eqnarray}
having defined the second-order electron-phonon matrix element in the Wannier representation:
  \begin{equation}
     g^{(2)}_{mn,\k\a,\k'\a'}(\bR_p,\bR_q,\bR_r) = \< w_{mq}| \frac{\partial^2 V_{\rm KS}}{\partial \tau_{\k\a p} \partial \tau_{\k'\a' 0}} | w_{nr}\>_{\rm sc}~.
  \end{equation} 
The inverse transformation of Eq.~\eqref{Eq:w2b_eph} is obtained by considering the orthonormality of the vibrational eigenmodes, the unitarity of the Wannier rotation matrices, and the wavevector sum rule $\sum_\bk \exp(i\bk\cdot\bR_p) = N_p \d_{p0}$~\cite{Giustino_2017}. The result it:
\begin{eqnarray}
\label{Eq:b2w_eph}
   &&   
    g^{(2)}_{mn,\k\a,\k'\a'}(\bR_p,\bR_q,\bR_r) = \nonumber \\ 
    && \hspace{10pt}= \frac{2}{N_p^4}
    \sum_{\nu\nu'} \sqrt{\frac{2M_\k\w_{\bq\nu}}{\hbar}} \sqrt{\frac{2M_{\k'}\w_{\bq'\nu'}}{\hbar}}
    \nonumber \\
    && \hspace{10pt}\times e^*_{\k\a \nu}e^*_{\k'\a' \nu'}  \sum_{m'n'} \red{U^\dagger_{mm' \bk+\bq+\bq'}} U_{n'n \bk}\nonumber\\
   && \hspace{10pt}\times \sum_{\bk,\bq,\bq'} e^{-i[\bk\cdot\bR_r+\bq\cdot\bR_p-(\bk+\bq+\bq')\cdot\bR_q]} 
   g^{(2)}_{m'n'\nu\nu'}(\bk,\bq,\bq')~.\nonumber \\
\end{eqnarray}
As in the case of Wannier-Fourier interpolation of the linear electron-phonon matrix elements~\cite{Giustino_Louie_2007}, these transformations can be used to interpolate second-order coupling \red{if $g^{(2)}_{mn,\k\a,\k'\a'}(\bR_p,\bR_q,\bR_r)$ is localized in real space. Preliminary calculations show that this is indeed the case,  see App.~\ref{App:locality}. The interpolation strategy is as follows}:
\begin{itemize}
    \item[(i)] Evaluate second-order matrix elements $g^{(2)}_{m'n'\nu\nu'}(\bk,\bq,\bq')$ on a coarse uniform Brillouin zone grid using the method described in Sec.~\ref{Sec:Derivations};
    \item[(ii)] Obtain MLWFs and their rotation matrices $U_{mn\bk}$ using standard methods, e.g., \textsc{Wannier90}~\cite{Mostofi_Marzari_2014};
    \item[(iii)] Use the results of the previous steps to obtain the second-order matrix elements $g^{(2)}_{mn,\k\a,\k'\a'}(\bR_p,\bR_q,\bR_r)$ via Eq.~\eqref{Eq:b2w_eph};
    \item[(iv)] Use the $g^{(2)}_{mn,\k\a,\k'\a'}(\bR_p,\bR_q,\bR_r)$ of the previous step inside Eq.~\eqref{Eq:w2b_eph} in order to compute 
    $g^{(2)}_{m'n'\nu\nu'}(\bk,\bq,\bq')$ at arbitrary wavevectors $\bk,\bq,\bq'$. This step relies on the fact that the matrix elements in the Wannier representation are usually short-ranged, so that a Fourier transform with zero padding is meaningful.
    This steps also requires the knowledge of the Wannier rotation matrices $U_{mn\bk}$ at the desired wavevectors; these matrices are obtained from standard Wannier interpolation of band structures~\cite{Marzari_Vanderbilt_2012}.
\end{itemize}
The procedure just described can be used as is for metallic systems. In the case of semiconductors and insulators, the matrix elements exhibit nonanalytic behavior at long wavelength. In the case of linear electron-phonon couplings, this complication is handled by removing the nonanalytic component, interpolating the short-ranged analytic part, and adding back the nonanalytic component at the end. This approach has successfully been demonstrated for the leading-order dipole singularity~\cite{Verdi_Giustino_2015,Sjakste_Mauri_2017} and for the next-to-leading order quadrupole singularity~\cite{Brunin_Hautier_2020, Park_Bernardi_2020, Jhalani_Bernardi_2020}. In the case of second-order electron-phonon couplings, the treatment of the dipole singularity has recently been developed in Ref.~\citenum{Houtput_Tempere_2025}; this development can directly be combined with the present method to yield full sampling of the second-order matrix elements across the Brillouin zone for polar insulators.

At the end of this section, we note that the same strategy leading to the interpolation formulas in Eqs.~\eqref{Eq:b2w_eph} and \eqref{Eq:w2b_eph} can be extended without difficulty to higher-order electron-phonon couplings if needed.

\section{Implementation and tests}\label{Sec:numerical}

\subsection{Workflow and computational setup}
\label{Sec:Computation}

The computational workflow is summarized in Fig.~\ref{Fig:Workflow}. We begin with a DFT calculation in the crystal \textit{unit cell} to obtain the self-consistent charge density. Then, using this density, we perform a non-self-consistent calculations of Kohn-Sham states and vibrational eigenmodes on a $N_1 \times N_2 \times N_3$ $\bk/\bq$ Brillouin-zone grid. 
In the next step, we proceed to a supercell calculation, and for this purpose we use a $N_1 \times N_2 \times N_3$ supercell. In the supercell, we slightly displace one atom, and we compute the self-consistent potential in the (distorted) ground state. Then the planewaves coefficients of the unit-cell wavefunctions $\psi_{n\bk}(\br)$ are read from file, and mapped into the corresponding coefficients of the supercell wavefunctions via Eq.~\eqref{Eq:corresp}. These two ingredients are used to evaluate the matrix elements in Eq.~\eqref{Eq:key_integral2}, and from these quantities we construct first- and second-order electron-phonon coupling matrix elements using Eq.~\eqref{Eq:fd_linear} and Eqs.~\eqref{Eq:2ndeph_simple}-\eqref{Eq:fd_diag}, respectively. 

In the following, we demonstrate these calculations by considering diamond, LiF, and graphite; these systems are chosen as examples of nonpolar semiconductors, polar semiconductors, and semimetals, respectively.

We perform DFT and DFPT calculations using the \textsc{Quantum Espresso} suite~\cite{Giannozzi_Baroni_2017}. For diamond and graphite, we employ the local density approximation (LDA) to the DFT exchange and correlation functional since it is well known to yield good lattice parameters. For LiF, we ue the generalized gradient approximation of Perdew, Burke, and Ernzerhof (PBE)~\cite{Perdew_Bruke_1996} for consistency with our prior work on polarons in this system~\cite{Dai_Giustino_2026}. In all cases, we use norm-conserving pseudopotentials~\cite{Hamann_2013, vanSetten_Rignanese_2018} and a planewaves kinetic energy cutoff of 100~Ry. The starting lattice vectors of all three systems are taken from the Materials Project database~\cite{Jain_Persson_2013}, and subsequently optimized; the atomic coordinates are fixed by their Wyckoff sites. 

To converge the electronic charge density in the unit cell, we choose $8\times8\times8$, $8\times8\times2$, and $6\times6\times6$ $\bk$-grids for diamond, graphite, and LiF, respectively. Unless otherwise specified, the unit-cell Kohn-Sham wavefunctions and phonon modes are computed on a $4\times4\times4$ $\bk/\bq$-grids in all cases; accordingly, we construct $4\times4\times4$ supercells. The displacement parameter $\delta$ in Eq.~\eqref{Eq:fd_linear} and Eqs.~\eqref{Eq:2ndeph_simple}-\eqref{Eq:fd_diag} is chosen by numerically taking the limit $\delta\rightarrow 0$, and is discussed in Sec.~\ref{Sec:Conv_dtau}.

\subsection{Displacement parameter for finite-difference formulas}
\label{Sec:Conv_dtau}

\begin{figure*}
    \centering
    \includegraphics[width=0.99\linewidth]{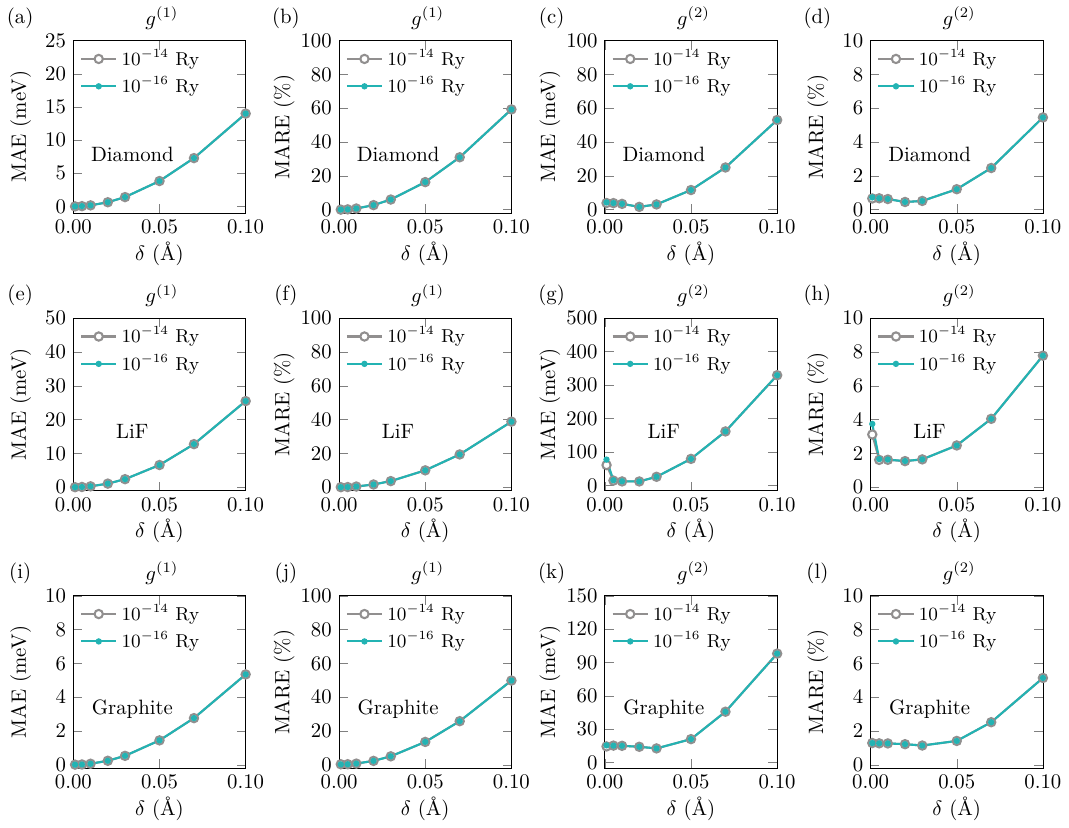}
    \caption{
    Benchmark of linear [$\gone$] and quadratic [$\gtwo$] electron-phonon coupling matrix elements computed via finite difference. The averages are evaluated over 4 valence bands and 4 conduction bands for diamond; 5 valence bands and 5 conduction bands for LiF; and 8 valence bands and 8 conduction bands for graphite.
    (a) and (b): Mean absolute error (MAE) and mean absolute relative error (MARE) of the linear electron-phonon coupling in diamond, respectively, plotted as a function of the atomic displacement amplitude $\d$ used in the finite-difference formulas. The ground truth corresponds to DFPT calculations of the matrix elements.
    (c) and (d): MAE and MARE of the quadratic electron-phonon coupling in diamond, respectively, plotted as a function of the atomic displacement amplitude. In this case, the ground truth is taken to be the Debye-Waller energy shift obtained from DFPT, as discussed in in Sec.~\ref{Sec:Conv_dtau}.
    The second rows of panels (e)-(h) show the same information as in (a)-(d), but this time for the polar insulator LiF. The third rows contains the same information as in (a)-(d), but this time for the semimetal graphite.    
    }
    \label{Fig:Conv_dtau}
\end{figure*}

In this section, we analyze the sensitivity of the first- and second-order electron-phonon matrix elements to the displacement parameter $\delta$ used in the finite-difference formulas given by Eqs.~\eqref{Eq:2ndeph_simple}-\eqref{Eq:fd_linear}.

Figures~\ref{Fig:Conv_dtau}(a) and (b) report the mean absolute error (MAE) and the mean absolute \red{relative} error (MARE) between the linear electron-phonon coupling matrix elements of diamond computed using (i) the present method for a given displacement parameter $\delta$, and (ii) using DFPT. The averages are evaluated over the $4\times 4\times 4$ Brillouin zone grid, 4 valence bands, and 4 conduction bands. We see that the MAE becomes extremely small and below 1~meV for $\delta = 0.01$~\AA. The corresponding MARE is of only 0.67\%. A similar trend is observed for LiF and graphite, which are shown in Fig.~\ref{Fig:Conv_dtau}(e)-(f) and (i)-(j), respectively: also in these cases, the MAE for $\delta = 0.01$~\AA\ is below 1~meV and the MARE are 0.69\% and 0.40\%, respectively. These tests validate our implementation and confirm that the present finite-difference approach using the resolution of identity in Eq.~\eqref{Eq:key_integral2} offers a level of accuracy comparable to standard DFPT for the linear electron-phonon coupling matrix elements.

Next, we move to the second-order matrix elements. Figures~\ref{Fig:Conv_dtau}(c) and (d) show the MAE and MARE for the second-order matrix elements of diamond, respectively, as a function of the displacement parameter $\delta$. Since there are no other methods for evaluating individual second-order matrix elements, for this test we compare the Debye-Waller energy shift given by Eq.~\eqref{Eq:dw_selfe}, which consists of a \textit{sum of diagonal} second-order matrix elements. This quantity can be computed using the Allen-Heine theory as implemented in \textsc{Quantum ESPRESSO} within the RIA~\cite{Lihm_Park_2020}; for consistency, in this test we also use the RIA version of our method. 
The figures show that the present method achieve an accuracy of 3~meV for $\delta = 0.01$~\AA, with a corresponding MARE of 0.63\%. Similar results are obtained for 
LiF and graphite in Figs.~\ref{Fig:Conv_dtau}(g)-(h) and (k)-(l), respectively, with MARE below 2\% in both cases. 

\begin{figure}
    \centering
    \includegraphics[width=0.99\linewidth]{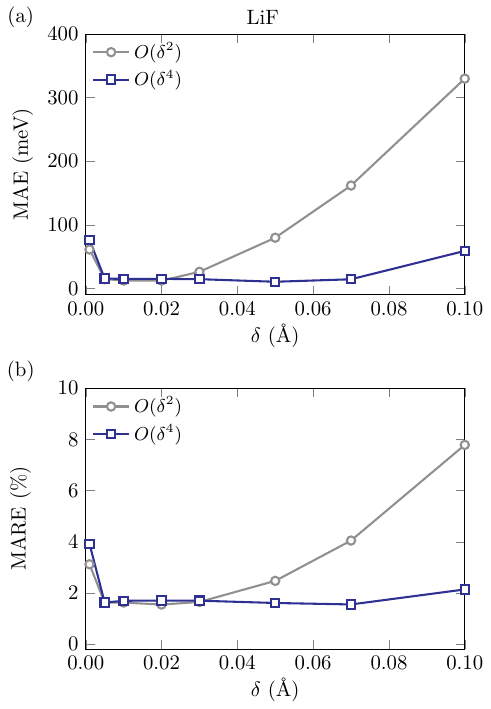}
    \caption{
    Dependence of the mean absolute error (a) and mean absolute relative error (b) on the finite-difference stencils for LiF. The fourth-order-accuracy stencil (blue curves) can reduce the sensitivity on $\d$ compared with second-order-accuracy stencil (gray curves).}
    \label{Fig:stencils}
\end{figure}

We note that the error tends to increase for $\delta < 0.02$~\AA. This observation is consistent with the fact that higher-order finite differences tend to accumulate numerical error at very small increments. To minimize these effects, we choose to proceed with $\d=0.02$~\AA\ in the following calculations. 
\red{One possible route to reduce the dependence on $\d$ is through the higher-order-accuracy finite-difference stencils, as can be seen in Fig.~\ref{Fig:stencils}, but this is at the cost significantly elevated computational costs and thus not pursued in this work.}

All the calculations shown in Figs.~\ref{Fig:Conv_dtau} were performed twice: first, using a convergence threshold for the total energy of $10^{-14}$~Ry in the BvK supercell; second, using a more stringent threshold of $10^{-16}$~Ry. We found that the resulting matrix elements do not change in any significant way, with the largest difference being 0.018~meV across all datasets with $\d = 0.02$~\AA\ .

\subsection{Validation of the rigid-ion approximation (RIA)}
\label{Sec:RIA_bad}

In this section, we investigate the accuracy of the RIA provided by Eq.~\eqref{Eq:2ndeph_RIA} by focusing first on diamond and then on LiF.

\subsubsection{RIA in diamond}\label{Sec:RIA_diam}

For this test, we compute the second-order electron-phonon matrix elements in two ways: (i) the complete matrix elements $g^{(2)}_{mn\nu\nu'}(\bk,\bq,\bq')$ using Eq.~\eqref{Eq:2ndeph_simple}; (ii) the RIA matrix elements $g^{(2),\text{RIA}}_{mn\nu\nu'}(\bk,\bq,\bq')$ using Eq.~\eqref{Eq:2ndeph_RIA}.

\begin{figure*}
    \centering
    \includegraphics[width=0.8\linewidth]{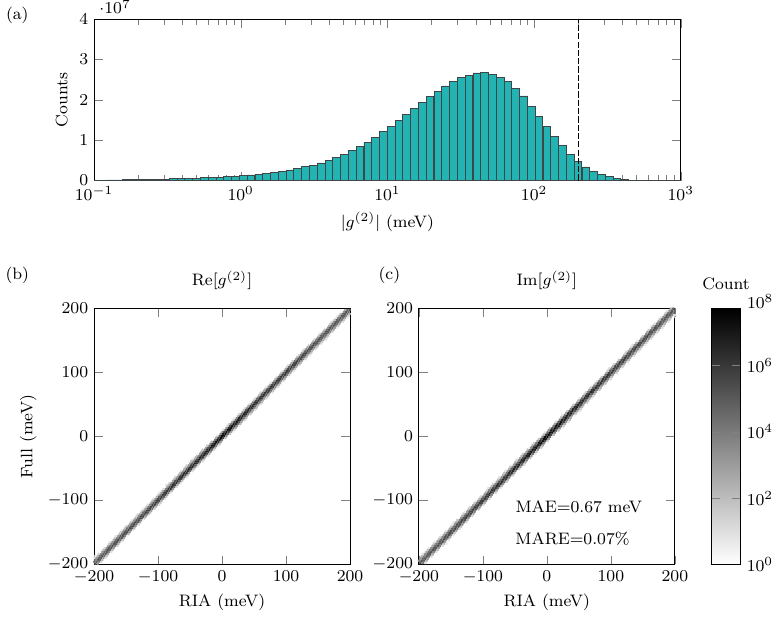}  
    \caption{Comparison of second-order electron-phonon coupling matrix elements [$\gtwo$] calculated under the RIA with the same quantity calculated by including all neighbor shells, for diamond. (a) The statistics of the absolute values of second-order electron-phonon matrix elements, showing that 98\% of these matrix elements are smaller than 200\,meV in magnitude, as indicated by the dashed line. (b) Comparison of the real part of $\gtwo$ between RIA and all-neighbor calculations, within the range of $\pm200$\,meV. To effectively present $6\times10^{8}$ datapoints, we binned the scatter plot via hexagonal tiles; darker tiles indicate more matrix elements within the tile. (c) Same as (b), but for the imaginary part of $\gtwo$. } 
    \label{Fig:RIA_vs_full}
\end{figure*}

Figure~\ref{Fig:RIA_vs_full}(a) shows the distribution of the second-order matrix elements calculated for diamond. While some matrix elements can be as large as 2~eV in magnitude, 98\% of these matrix elements have magnitude below 200~meV as indicated by the vertical line in the figure. 
Keeping this in mind, in Figs.~\ref{Fig:RIA_vs_full}(b) and (c) we show a comparison between the full matrix elements and the RIA matrix elements in the range $\pm 200$~meV.
We see that, while the full matrix elements and their RIA counterparts are not identical, they are very close throughout the entire range, with no outliers. In particular, the MAE is as small as 0.67~meV, and the MARE is 0.07\%. This comparison suggests that the RIA is a good approximation for calculating second-order electron-phonon matrix elements in diamond.

\subsubsection{RIA in LiF}\label{Sec:RIA_LiF}

In Sec.~\ref{Sec:RIA_diam} we tested the RIA on diamond by considering the Debye-Waller energy shift. Since this is a cumulative metric that involves a sum over all phonons, we now test the RIA in LiF by considering a \textit{single} vibrational mode.

To this end, we distort the LiF crystal by displacing atoms along a single vibrational mode $(\bq\nu)$, and we calculate the corresponding change of the Kohn-Sham eigenvalue of the valence band maximum (VBM). For this calculation, we employ the second-order polaron equations derived in Eq.~\eqref{Eq:2nd_plrneq_1_rec} and use the displacement coefficient $B_{\bq \nu}$ as a tunable parameter. We emphasize that, in this test, we are not solving the coupled polaron equations Eqs.~\eqref{Eq:2nd_plrneq_1_rec} and \eqref{Eq:2nd_plrneq_2_rec}; instead, we calculate the Kohn-Sham eigenvalues by solving the first of these equations at \textit{fixed} crystal structure, without polarons.

\begin{figure}
    \centering
    \includegraphics[width=0.99\linewidth]{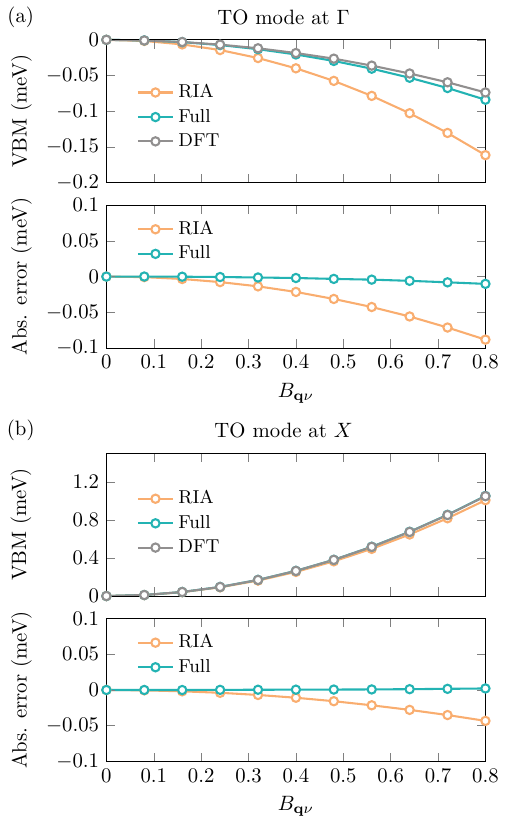}
    \caption{
    Testing the accuracy of the RIA in LiF. (a) \red{Upper pannel:} The VBM energy as a function of the collective, dimensionless displacement coordinate $B_{\bq\nu}$, corresponding to the TO mode at $\Gamma$. 
    \red{Lower panel: The absolute error of RIA and full-accuracy calculations that include all neighbors (``full'').}
    In this example, the RIA (orange curve) yields a nearly 100\% \red{relative} error as compared to direct supercell DFT calculations (gray curve). On the other hand, the calculations including all neighbors (blue curve)  correctly predict the VBM shift. (b) Same as (a), but for the highest TO mode at the $X$ point [$\bq=(0.0, 0.0, 0.5)$ in crystal coordinates]. In this case, both the RIA and all-neighbor calculations yield similar results and agree with direct supercell DFT calculations.}
    \label{Fig:Bad_RIA}
\end{figure}

Figure~\ref{Fig:Bad_RIA}(a) shows the comparison between (i) a direct supercell calculation, (ii) a calculation using Eq.~\eqref{Eq:2nd_plrneq_1_rec} and the full matrix element $g^{(2)}_{mn\nu\nu'}(\bk,\bq,\bq')$, and (iii) a calculation using Eq.~\eqref{Eq:2nd_plrneq_1_rec} and the RIA matrix element $g^{(2),\text{RIA}}_{mn\nu\nu'}(\bk,\bq,\bq')$, for the zone-center optical phonon at $\Gamma$. From this plot we see that the full second-order matrix element captures almost exactly the result of the DFT supercell calculation, for a collective displacement of all lithium atoms up to 0.016~\AA, corresponding to $B_{\bq\nu}=0.8$ in Fig.~\ref{Fig:Bad_RIA}(a). On the other hand, the RIA severely overestimates the change in the VBM energy, by approximately a factor of 2 throughout the entire range of displacements. 

On the other hand, when we compare these approaches for the highest transverse optical (TO) mode at the $X$ point in
Fig.~\ref{Fig:Bad_RIA}(b) [$\bq=(0,0,0.5)$ in crystal coordinates], we see that both the full matrix element and the RIA capture supercell results very accurately throughout the entire range of displacements. 

These findings suggest that, while the RIA appears \red{to yield small absolute errors [lower panels of Fig.~\ref{Fig:Bad_RIA}(a) and \ref{Fig:Bad_RIA}(b)] and} perform well when sums over matrix elements are computed, its reliability is less clear-cut when one is interested in properties that depend on individual matrix elements at specific wavevectors. This may be the case, for example, in the study of Kohn anomalies~\cite{Caruso_Giustino_2017, Pisana_Mauri_2007}, conventional superconductors where specific phonons drive pairing~\cite{Liu_Kortus_2001, An_Pickett_2001, Kortus_Boyer_2001,Giustino_Louie_2007b, Blase_Connetable_2004, Boeri_Andersen_2004}, light-driven coherent phonons~\cite{Thouin_Kandada_2019, Caruso_Zacharias_2023, Emeis_Caruso_2025, Pradeep_Kanatzidis_2026}, and phonon-induced renormalization of the total energy~\cite{Ponce_Gonze_2025}. In all these cases, one should always aim to perform validation tests like those discussed in this section before drawing physical conclusions from RIA calculations.

\subsection{Validation of the finite-range approximation (FRA)}
\label{Sec:LongerRange}

\begin{figure*}
    \centering
    \includegraphics[width=0.9\linewidth]{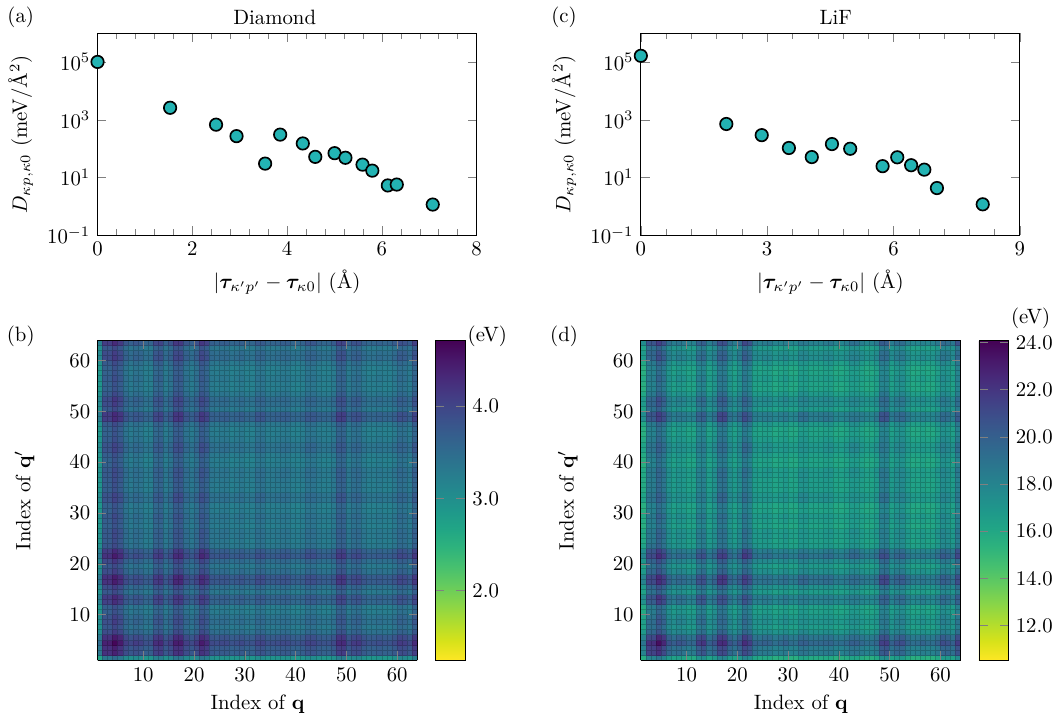}  
    \caption{
    (a) Range of second-order electron-phonon matrix elements in real space. The scatter plot shows  the quantity $D_{\k' p', \k 0}$ as defined in Eq.~\eqref{Eq:metric_D} as a function of the distance between the pair of atoms $|\btau_{\k' p'}-\btau_{\k 0}|$, for diamond. \red{The data points with the same $|\btau_{\k' p'}-\btau_{\k 0}|$ have been added up so that each distance corresponds to only one data point in this plot.}
    (b) Color map showing the magnitude of $D_{\bq,\bq'}$ as defined by Eq.~\eqref{Eq:metric_Dq}, for diamond. The $\bq\ne \bq'$ components of these matrix elements can be as large as the diagonal elements. The $\bq$-point index is defined by $n = (i-1)N^2 + (j-1)N + (k-1) + 1$, with $i,j,k$ ranging between 1 and $N=4$ and the $\bq$-point coordinates being $(i-1,j-1,k-1)/N$ in crystal coordinates. \red{(c) and (d) The same as (a) and (b), respectively, but for LiF.}}
    \label{Fig:Matrix_amp}
\end{figure*}

In this section we go beyond the RIA and explore the FRA outlined in Sec.~\ref{Sec:RIA_better}.
Figure~\ref{Fig:Matrix_amp}(a) \red{and \ref{Fig:Matrix_amp}(c)} show how the second-order derivatives of the Kohn-Sham potential [Eq.~\eqref{Eq:RIA-decomp}] of diamond and LiF, respectively, decay as a function of the distance between atoms. In order to provide this information in a more compact form, we plot the quantity: 
\begin{align}
    \label{Eq:metric_D}
    D_{\k' p', \k 0} 
    \hspace{-0.1cm}
    =
    \hspace{-0.1cm}
    \left[\frac{1}{N_p^2}\! \sum_{mn,\bk\bk'}\sum_{\a\a'}
    |{\bra{m\bk'} \frac{\D^2 \hat{V}_{\rm KS}}{\D \tau_{\k' \a' p'} \D \tau_{\k \a 0}} \ket{n\bk} }|^2\right]^{\!1/2}\!\!\!\!\!\!.
\end{align}
The band average is performed over 4 valence bands, 4 conduction bands, and a $4\times 4\times 4$ gird in the Brillouin zone. As defined, this quantity depends only on the atomic coordinates and is positive definite. \red{In Fig.~\ref{Fig:Matrix_amp}(a) and \ref{Fig:Matrix_amp}(c), we further add up $D_{\k' p', \k 0}$ with the same interatomic distance so that each distance corresponds to only one data point.} The plot of $D_{\k' p', \k 0}$ vs.\ the interatomic distance $|\btau_{\k' p'}-\btau_{\k 0}|$ shows that the second derivatives are very short-ranged in the case of diamond. In fact, $D_{\k' p', \k 0}$ for nearest-neighbor atoms is already two orders of magnitudes smaller than the derivatives taken with respect to the same atom.

It may be worth noting that the short-ranged nature of $D_{\k' p', \k 0}$ does not imply that the off-diagonal second-order electron-phonon matrix elements can be neglected. To illustrate this point, in Fig.~\ref{Fig:Matrix_amp}(b) and \red{\ref{Fig:Matrix_amp}(d)}, we plot the following average of the matrix elements over the electron bands and electron wavevector:
\begin{align}
    \label{Eq:metric_Dq}
    D_{\bq,\bq'} =
    \left[\frac{1}{N_p}\! \sum_{mn\bk} \sum_{\nu \nu'}
    |g^{(2)}_{mn\nu\nu'}(\bk,\bq,-\bq')|^2\right]^{\!1/2}\!\!\!\!\!\!.
\end{align}
This analysis confirms indeed that off-diagonal second-order couplings with $\bq\ne\bq'$ can be as large as the corresponding diagonal couplings, therefore they cannot \textit{a priori} be neglected, \red{which would otherwise yield unphysical calculation results in, for example, polaron calculations, which depend explicitly on these off-diagonal matrix elements [Eqs.~\eqref{Eq:2nd_plrneq_1_rec} and \eqref{Eq:2nd_plrneq_2_rec}].} 

\begin{figure*}
    \centering
    \includegraphics[width=0.95\textwidth]{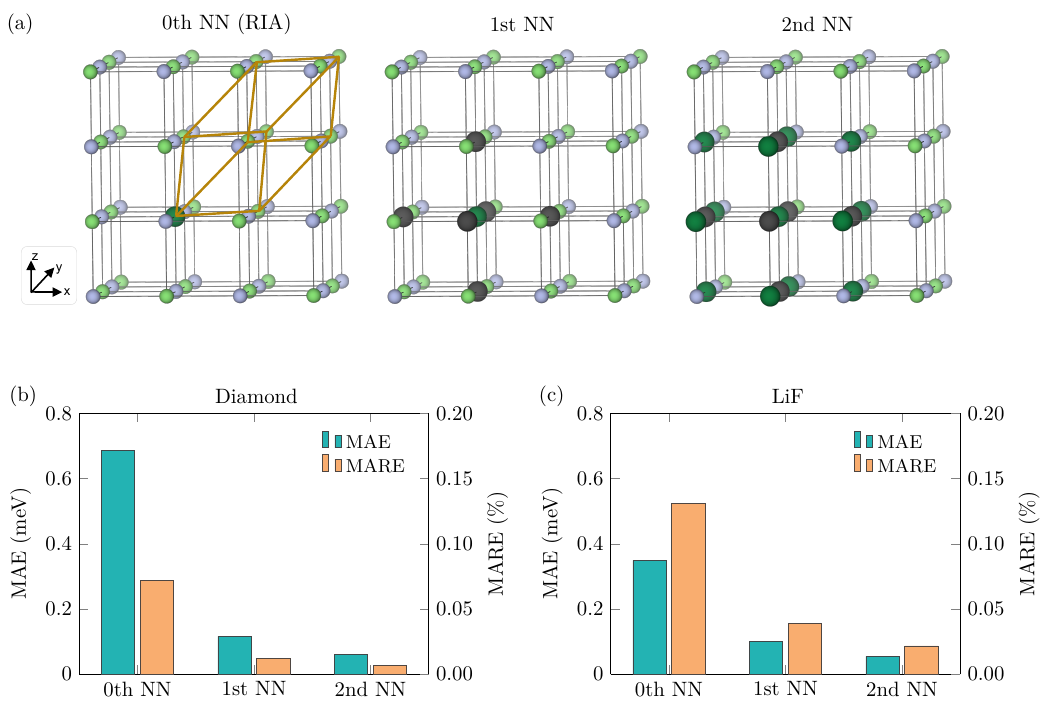}
    \caption{
    Hierarchy of approximations for calculating second-order electron-phonon coupling matrix elements beyond the RIA. (a) Schematic illustration of different approximations. Here, the Li atom of LiF in the reference primitive cell is chosen as an illustrative example. Green is for Li, gray is for F.
    The orange polyhedron indicates the primitive unit cell of LiF. The atoms that are included in the mixed derivatives are indicated by the enlarged radius and darkened color. 
    (b) MAE (cyan bars, left vertical axis) and MARE (orange bars, right vertical axis) of the matrix elements calculated using the ranges shown in (a), compared with the all-neighbor matrix elements, for diamond. (c) Same layout as in (b), this time for LiF. 
    }
    \label{Fig:Diff_approxs}
\end{figure*}

In Fig.~\ref{Fig:Diff_approxs} we analyze the accuracy of the RIA (0-th NN approximation) as well as the lowest FRAs corresponding to retaining the 1-st NN  shelland the 2-nd NN shell in Eq.~\eqref{Eq:2ndeph_simple_FRA}, respectively. The MAE and the MARE are evaluated with respect to calculations performed by including all pairs of atoms. In the case of diamond [Fig.~\ref{Fig:Diff_approxs}(b)] we see that the errors decrease significantly with respect to the RIA, by approximately $6\times$, already when we include the 1-st NN shell. The relative improvement upon adding the 2-nd NN shell is a more modest factor $2\times$. Similar observations can be made for LiF [Fig.~\ref{Fig:Diff_approxs}(c)], where the error decreases $3\times$ from the RIA to the 1-st NN shell FRA, and then only marginally upon including the 2-nd NN shell. 

Since the number of required supercell calculations scales with the number of atoms included in the FRA, and since the accuracy gain becomes marginal beyond the 1-st NN shell, the 1-st NN shell FRA appears to be the optimal choice in terms of balancing accuracy and computational cost. We note that, with this choice, the MAE is approximately 0.1~meV for both diamond and LiF, and the corresponding MARE is about 0.01\% in both cases.

\section{Application to polarons}
\label{Sec:Application}

\begin{table}[b]
\begin{tabular}{l@{\hspace{10pt}}c@{\hspace{10pt}}c@{\hspace{10pt}}}
\toprule
   & Supercell & Linear Pol. Eqs. \\ 
\midrule
Infinite supercell & & \\
\quad Formation energy (meV) &1702 & 1980 \\
\quad Excitation energy (meV) & 4176 & 4760\\
\midrule
3$\times$3$\times$3 supercell & & \\
\quad Formation energy (meV) & 652  & 887 \\
\quad Excitation energy (meV) & 2106 & 2487 \\
\bottomrule
\end{tabular}
\caption{Comparison between the energetics of the one-center, small hole polaron in \ch{LiF}, as obtained from supercell DFT calculations \cite{Sadigh_Aberg_2015} and from the linear polaron equations~\cite{Sio_Giustino_2019a}. The linear polaron equations tend to overestimate the supercell values, both in the limit of infinite supercell and in the case of the small 3$\times$3$\times$3 supercell. For convenience, in this work we focus on the small supercell.}
\label{Tab:eform}
\end{table}

In this section we investigate the impact of the second-order electron-phonon matrix elements, Eq.~\eqref{Eq:2ndeph_def}, in the calculation of polarons using the polaron equations discussed in Sec.~\ref{Sec:Polaroneq}.

\begin{figure}
    \centering
    \includegraphics[width=0.7\linewidth]{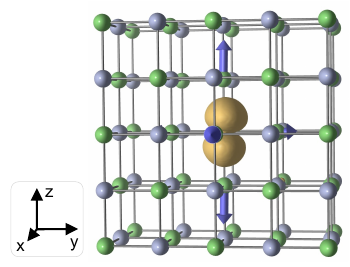}
    \caption{One-center, small hole polaron in LiF, with Li in green and F in grey. The yellow isosurface is the polaron charge density obtained from the direct supercell DFT calculation, and the blue arrows indicate the corresponding atomic displacements with respect to the undistorted crystal. }
    \label{Fig:LiF_illu}
\end{figure}

For this test we focus on the small, one-center hole polaron in LiF, which is shown in Fig.~\ref{Fig:LiF_illu}. This system was investigated in a number of previous studies~\cite{Sio_Giustino_2019a, Sio_Giustino_2019b, Vasilchenko_Gonze_2025, Robinson_Reichman_2025, Lee_Bernardi_2021, Luo_Bernardi_2022, Luo_Bernardi_2025, Falletta_Pasquarello_2025}. In Ref.~\citenum{Dai_Giustino_2024b}, we showed that the linear polaron equations [Eqs.~\eqref{Eq:2nd_plrneq_1_rec},\eqref{Eq:2nd_plrneq_2_rec} with $g^{(2)}_{mn\nu}(\bk,\bq,\bq')$ set to zero] are unable to fully capture the energy landscape of this polaron that one obtains from explicit supercell calculations~\cite{Dai_Giustino_2026}. As a baseline for the following discussion, in Table~\ref{Tab:eform} we summarize these prior findings. In Ref.~\citenum{Dai_Giustino_2024b}, we also hypothesized that the discrepancy must arise from the very large atomic displacements associated with this polaron, which reach up to 0.4~\AA. Our hypothesis was that, for such large displacements, linear electron-phonon couplings may not be able to correctly describe polaron energetics. Here, we test this hypothesis by solving the quadratic polaron equations. We will see that the discrepancy is indeed resolved once second-order electron-phonon couplings are taken into account.

For these calculations we solve the polaron equations on a $3\times3\times3$ Brillouin zone grid, and we compare the results with explicit DFT calculations in a equivalent $3\times3\times3$ BvK supercell~\cite{Sadigh_Aberg_2015}. In this comparison, we \textit{do not} include finite-size corrections, that is we do not seek to obtain the energy landscape in the limit of infinite supercell size. The changes in polaron energies from the $3\times3\times3$ supercell to the extended limit are reported in Tab.~\ref{Tab:eform}. The choice of using a small BvK supercell allows us to perform extensive convergence tests with respect to the number of bands included in the polaron equations, as discussed in App.~\ref{App:band_conv}.

Based on the analysis of the FRA in Sec.~\ref{Sec:LongerRange}, we opt to compute second-order matrix elements within the 1-st NN shell FRA, which yields a MAE as small as 0.1~meV as compared to calculations including all possible pairs of atoms in the mixed second-order derivatives.

\begin{figure*}
    \centering
    \includegraphics[width=0.99\linewidth]{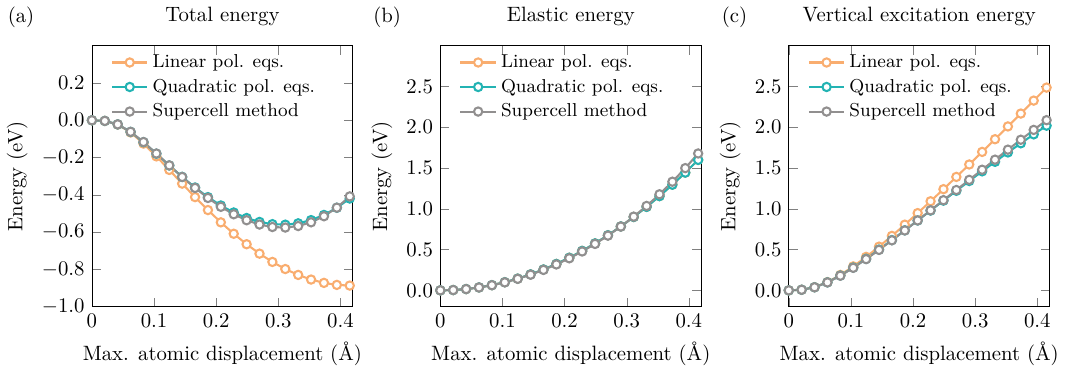}
    \caption{
    (a) Potential energy surface of the one-center, small hole polaron in LiF, as computed within the supercell approach (gray), the linear polaron equations (orange), and the second-order polaron equations (green). The atomic coordinates are obtained from linear interpolation between the pristine LiF crystal (no displacement) and the configuration of the hole polaron obtained from the linear polaron equations (maximum atomic displacement $\simeq$0.4\AA). The zero of the energy corresponds to the delocalized solution (i.e., no polaron).
    (b) and (c): Decomposition of the potential energy surface shown in (a) into elastic energy and electronic excitation energy, respectively.}
    \label{Fig:Energies}
\end{figure*}

\subsection{Polaron formation energy}\label{Sec:Polaroneq_sol}

Figure~\ref{Fig:Energies}(a) shows the potential energy landscape calculated for the hole polaron in LiF using three distinct approaches: (i) a direct supercell calculation including self-interaction correction~\cite{Dai_Giustino_2024a,Sadigh_Aberg_2015}; (ii) a calculation using the polaron equations with only first-order electron-phonon couplings, i.e., Eqs.~\eqref{Eq:2nd_plrneq_1_rec},\eqref{Eq:2nd_plrneq_2_rec} with $g^{(2)}_{mn\nu}(\bk,\bq,\bq')$ set to zero; (iii) a calculation where the second-order matrix elements are included. 

In order to compare these three methods along a single collective coordinate, we first determine the polaron structure using the linear polaron equations, and then we construct intermediate structures between the pristine crystal and the polaron configuration by linear interpolation. With this choice, the structure at zero displacement in Fig.~\ref{Fig:Energies}(a) corresponds to the undistorted crystal, while the structure with a ``maximum displacement'' of 0.4~\AA\  is the polaron reported in Ref.~\cite{Dai_Giustino_2026} and shown in Fig.~\ref{Fig:LiF_illu}. For each intermediate configuration, the polaron equations yield the formation energy:
\begin{align}
    \label{Eq:Etot_plrneq}
    \Delta E_{\rm f}
    =
    -(\ve_{\rm p} - \ve_{\rm VBM})
    +
    \frac{1}{N_p} \sum_{\bq \nu} |B_{\bq \nu}|^2 \hbar \w_{\bq \nu}, 
\end{align}
where $\Delta E_{\rm f}$ is referred to the undistorted structure, and $\ve_{\rm VBM}$ is the VBM of the same undistorted structure. The polaron eigenvalue $\ve_{\rm p}$ is computed by solving Eq.~\eqref{Eq:2nd_plrneq_1_rec} using the coefficients $B_{\bq \nu}$ obtained by inverting Eq.~\eqref{Eq:dtau_decom}. In Eq.~\eqref{Eq:Etot_plrneq}, the first term on the right-hand side corresponds to the vertical excitation energy, and the second term corresponds to the elastic energy~\cite{Dai_Giustino_2026}. 

Within the direct supercell approach, the formation energy is calculated via:
\begin{align}   
    \label{Eq:Etot_DFT}
    \Delta E_{\rm f}
    =
    -(\ve_{\rm p} - \ve_{\rm VBM})
    + [E^N(\btau) - E^N(\btau_0)]~.
\end{align}
As in Eq.~\eqref{Eq:Etot_plrneq}, the first and second terms on the right-hand side of these expression correspond to the vertical excitation energy and the elastic energy, respectively~\cite{Dai_Giustino_2026}. 

In Fig.~\ref{Fig:Energies}(a) we see that the formation energy obtained from linear polaron equations (orange curve) overestimates the direct supercell result (gray curve) by as much as 0.4~eV, which corresponds to about 53\% of the formation energy. This overestimation yields a potential well that is too deep and also shifted toward too large atomic displacements. The deviation between linear polaron equations and supercell calculations starts becoming significant for atomic displacements exceeding 0.2~\AA.

Upon including the second-order electron-phonon matrix elements $g^{(2)}_{mn\nu}(\bk,\bq,\bq')$ in Eq.~\eqref{Eq:2nd_plrneq_1_rec} (blue curve), we see that the discrepancy is almost entirely removed, the potential well becomes shallower, and the minimum is found at smaller atomic displacements. This result is fully consistent with the direct supercell approach, the remaining residual error being smaller than 11~meV through the entire range of coordinates. This finding indicates that second-order electron-phonon couplings are essential to describe strong-coupling polarons as found in LiF.

To further confirm this point, in Fig.~\ref{Fig:Energies}(b) and (c) we decompose the formation energy into the elastic energy and the vertical excitation energy, respectively. In panel (b), we see that the elastic energy is almost identical across the three methods considered here. This observation indicates that anharmonic effects play a negligible role in this case; to avoid confusion, we emphasize that by ``anharmonic effects'' we mean cubic and higher-order corrections to the total energy of the ground-state structure with $N$ electrons. Such contributions are included by construction in the direct supercell approach, but are absent in our calculations using the polaron equations. On the other hand, in panel (c) we see that the vertical excitation energy is strongly affected by the inclusion of second-order electron-phonon couplings. This indicates that (i) the discrepancy between the linear polaron equations and the supercell approach originates from neglecting the second-order change of the polaron eigenvalue with respect to the atomic displacements; and (ii) while second-order matrix elements are essential, the contributions of third- and higher-order corrections are negligible.

\subsection{Polaron hopping barrier}
\label{Sec:Hopping}

\begin{figure}
    \centering
    \includegraphics[width=0.99\linewidth]{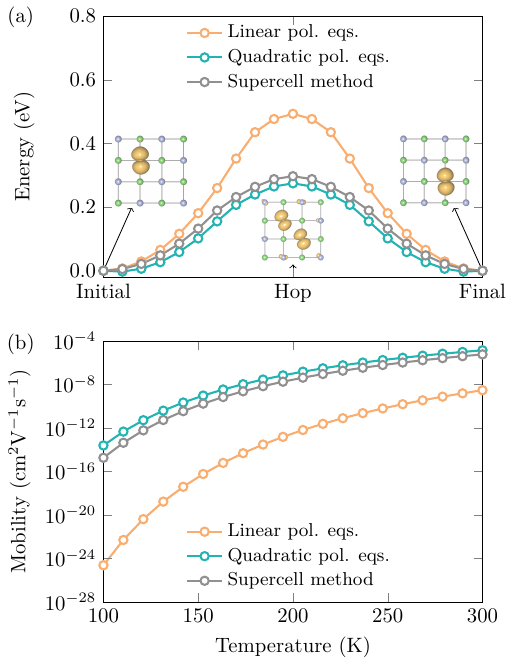}  
    \caption{(a) Hopping barrier of the one-center, small hole polaron in LiF, as calculated via the supercell approach (gray), the linear polaron equations (orange), and the \red{second-order} polaron equations (green). The atomic coordinates on the horizontal axis are obtained by linearly interpolating between the initial and final structure shown as insets.
    (b) Estimate of polaron mobility in LiF using the Emin–Holstein–Austin–Mott theory~\cite{Deskins_Dupuis_2007}. The plots are generated by extracting the polaron hopping barrier from panel (a), and the color code is the same.}
    \label{Fig:Hopping}
\end{figure}

Since, in the case of LiF, second-order electron-phonon couplings can modify the polaron formation energy by nearly a factor $2\times$, it is possible that the same corrections may modify the polaron hopping barrier and impact polaron transport properties.

To check this point, we compute the hopping barrier of the small hole polaron in LiF, $\Delta E_{\rm hop}$, and we estimate the polaron mobility via the Emin-Holstein-Austin-Mott theory ~\cite{Deskins_Dupuis_2007, Morita_Walsh_2023}.

To evaluate $\Delta E_{\rm hop}$, we consider a hopping path starting with a hole polaron localized on a F atom and ending with the same polaron localized on a nearest-neighbor F atom.  
These initial and final states are shown in Fig.~\ref{Fig:Hopping}(a), and the intermediate configurations are generated by linear interpolation~\cite{Morita_Walsh_2023}. More refined approaches such as the nudged elastic bands could be used~\cite{Henkelman_Jonsson_2000}, but the substance of the following analysis would not change. The polaron energies along this path, using either the polaron equations or the supercell approach, are evaluated in the same way as in Sec.~\ref{Sec:Polaroneq_sol}. From the calculated barrier $\Delta E_{\rm hop}$, we estimate the mobility using the standard expression: 
\begin{equation}
    \mu = \frac{e\,R^2 \w}{2\pi k_{\rm B} T}   \exp(-\Delta E_{\mathrm{hop}}/k_{\rm B} T) ~,\label{Eq:mobility}
\end{equation}
where $R$ is the distance between the initial and final sites ($R=2.87$~\AA); $\w$ is the characteristic frequency of the optical phonons that mediate intersite hopping, which we set to the highest-energy optical modes ($\hbar\w = 75$~meV, Ref.~\citenum{Sio_Giustino_2019b}); $k_{\rm B}$ is the Boltzmann constant, and $T$ is the absolute temperature. We emphasize that Eq.~\eqref{Eq:mobility} provides only an estimate for the mobility, and more sophisticated approaches are needed for quantitative predictions~\cite{Birschitzky_Franchini_2025, Wu_Ping_2018}.

Figure~\ref{Fig:Hopping}(a) shows that including only linear electron-phonon interactions (orange curve) overestimates the hopping barrier in LiF by approximately a factor of two compared to direct supercell calculations (gray curve). On the other hand, when second-order electron-phonon interactions are included (blue curve), we find that the resulting hopping barrier closely matches the direct supercell result.

Owing to the exponential dependence of the mobility on $\Delta E_{\rm hop}$ in Eq.~\eqref{Eq:mobility}, the fact that linear electron-phonon couplings overestimate the barrier leads to a severe underestimation of the polaron mobility, by many orders of magnitude, as seen in Fig.~\ref{Fig:Hopping}(b). Also in this case, once second-order electron-phonon couplings are taken into account, the correct order of magnitude of the polaron mobility is recovered.

At the end of this section it is worth noting that the small hole polaron in LiF likely represents an extreme case of strong-coupling polaron due to its large formation energy and colossal atomic displacements. Nevertheless, the present analysis clearly demonstrates that nonlinear electron-phonon interactions play an essential role in small-polaron transport, and must be accounted for when using electron-phonon Hamiltonians to describe polaron energetics and dynamics. 

\section{Conclusions and outlook}
\label{Sec:Conclusion}

In conclusion, we developed a systematic and efficient first-principles framework for computing nonlinear electron-phonon coupling matrix elements.
The method is based on finite differences and avoids the need for higher-order DFPT calculations. Using second-order electron-phonon interactions as a representative case, we showed that our approach not only reproduces previous results based on the RIA, but also provides the complete second-order coupling matrix elements with controlled accuracy. Furthermore, by incorporating these matrix elements into the \textit{ab initio} polaron equations, we demonstrate the critical role of nonlinear electron-phonon couplings in the physics of small polarons. 

This work also opens several new directions for future research. First, the formal simplicity of the present finite-difference approach makes this method naturally compatible with hybrid-functional calculations as well as GW calculations, thereby enabling calculations of both linear and nonlinear electron-phonon interactions beyond standard DFT. This is especially important since it is well known that DFT electron-phonon coupling matrix elements tend to be somewhat underestimated, and post-DFT corrections such as GW perturbation theory (GWPT)~\cite{Li_Louie_2019} must be performed to achieve predictive accuracy. Beyond nonlinear electron-phonon couplings, we emphasize that the present development also provides an alternative, efficient, and accurate approach for computing linear electron-phonon couplings beyond DFT. 

Second, as discussed in Sec.~\ref{Sec:Wannier}, nonlinear electron-phonon matrix elements can efficiently be interpolated using a generalization of the Wannier-Fourier interpolation of Ref.~\citenum{Giustino_Louie_2007}. While this extension is straightforward for metals, insulators present additional challenges due to long-range electrostatic interactions. In particular, dipolar and quadrupolar fields associated with atomic vibrations induce non-analytic behaviors of the coupling matrix elements in the long-wavelength limit~\cite{Verdi_Giustino_2015, Park_Bernardi_2020, Brunin_Hautier_2020}. A detailed investigation of these long-range contributions will be essential for developing accurate interpolation schemes; the treatment of second-order Fr\"ohlich coupling of Ref.~\citenum{Houtput_Tempere_2025} constitutes an excellent starting point in this direction.

Third, the present method is well suited to investigate electron-phonon physics in strongly anharmonic materials, from halide perovskites to superconducting hydrides, where the harmonic approximation for lattice dynamics breaks down and large atomic displacements are common. In such systems, electron-phonon interactions can extend well beyond the linear regime, yet most existing studies still rely on linear coupling combined with effective harmonic models~\cite{Zacharias_Even_2026,Zacharias_Even_2023a,Bianco_Errea_2023, Monacelli_Mauri_2021, Knoop_Hellman_2024}. The present approach enables systematic investigations of second- and higher-order electron-phonon interactions in these highly anharmonic materials.

Finally, nonlinear electron-phonon interactions are expected to play a role in a broad range of phenomena that have traditionally been described using linear electron-phonon coupling only, including charge transport, phonon-mediated optical processes, excited-state dynamics, and superconductivity. While the importance of nonlinear couplings is increasingly recognized in the context of effective Hamiltonians~\cite{Zappacosta_Tempere_2025, Klimin_Mishchenk0_2024, Houtput_Franchini_2026}, fully \textit{ab initio} investigations remain scarce. We hope that this work will make it possible to address these challenges and to investige new aspects of electron-phonon physics and its manifestations.

\vspace{5pt}
\noindent
\textbf{Data and Code availability}\\
\red{The code of calculating nonlinear electron-phonon coupling matrix elements is implemented within \textsc{EPW} under the branch \texttt{fdeph}} at \href{https://gitlab.com/epw/q-e}{https://gitlab.com/epw/q-e}.

\begin{acknowledgments}
We are grateful to Jon Lafuente-Bartolom\'e, Donghwan Kim, and Kaifa Luo for insightful discussions. This research was supported by the Computational Materials Sciences Program funded by the US Department of Energy, Office of Science, Basic Energy Sciences, under award no. DE-SC0020129. This research used resources of the National Energy Research Scientific Computing Center and the Argonne Leadership Computing Facility, which are Department of Energy Office of Science User Facilities supported by the Office of Science of the US Department of Energy, under Contract Nos. DE-AC02-05CH11231 and DE-AC02-06CH11357, respectively. We also acknowledge the Texas Advanced Computing Center at The University of Texas at Austin for providing access to Frontera, Stampede3, and Lonestar6 (http://www.tacc.utexas.edu).
\end{acknowledgments}
\appendix

\section{\red{Locality of the second-order electron-phonon coupling in real space}}
\label{App:locality}

\begin{figure}
    \centering
    \includegraphics[width=0.99\linewidth]{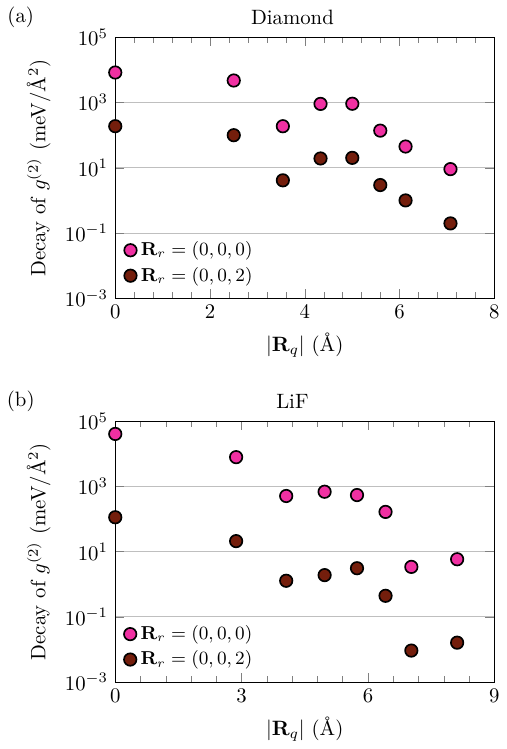}
    \caption{\red{Decay of the second-order electron-phonon coupling matrix elements in real space [defined in Eq.~\eqref{Eq:decay}] with respect to electronic degrees of freedom, $\bR_q$ and $\bR_r$, for diamond (a) and LiF (b). $\bR_p$, which corresponds to the nucleus degree of freedom, is fixed to be $(0, 0, 0)$ in the unit of primitve-cell lattice vectors, i.e. the central unit cell in the BvK supercell. In both cases, the real-space $g^{(2)}$ decay exponentially with increasing $\bR_q$, and when $\bR_r$ is set to be away from the central unit cell (brown discs), Eq.~\eqref{Eq:decay} will be orders of smaller than that where $\bR_r$ is set to coincide with $\bR_p$ (pink discs).}
    }
    \label{Fig:locality}
\end{figure}

\red{To test the feasibility of Wannier-interpolating the second-order electron-phonon coupling matrix elements, we have calculated $g_{mn,\k\a,\k'\a'}^{(2)}(\bR_p, \bR_q, \bR_r)$, which is defined in Eq.~\eqref{Eq:b2w_eph}, for diamond and LiF and investigate how they decay with respect to $\bR_q$ and $\bR_r$~\cite{Marzari_Vanderbilt_2012}, in addition to the decay with respect to $\bR_p$ that is already shown to be fast in Fig.~\ref{Fig:Matrix_amp}. 
To this end, we calculated the quantity
\begin{align}
    \label{Eq:decay}
    \left( 
    \sum_{mn}|g^{(2)}_{mn,\k\a,\k'\a'}(\bR_p,\bR_q, \bR_r)|^2
    \right)^{1/2},
\end{align}
and one representative result can be found in Fig.~\ref{Fig:locality}, where we fix $\bR_p$ to be the central unit cell in the BvK supercell, and choose $\k\a$ to correspond to one carbon atom and fluorine atom in diamond and LiF moving along x direction, respectively. In addition, we set $\k\a=\k'\a'$, which will yield the largest magnitude of $g_{mn,\k\a,\k'\a'}^{(2)}(\bR_p, \bR_q, \bR_r)$ with respect to degrees of freedom of nuclei, as has been demonstrated in Fig.~\ref{Fig:Matrix_amp}}. 

\red{The calculation results reveal that, for both diamond and LiF, the quantity defined in Eq.~\eqref{Eq:decay} will decay exponentially with respect to $\bR_q$ irrespective of $\bR_r$. Furthermore, when $\bR_r$ is chosen to be $(0,0,2)$ (brown discs in Fig.~\ref{Fig:locality}), which is away from $\bR_p$ [$\bR_p = (0,0,0)$ in the unit of primitive lattice vectors], Eq.~\eqref{Eq:decay} is smaller by orders of magnitude compared to the situation where $\bR_r$ coincides with $\bR_p$ (pink discs in Fig.~\ref{Fig:locality}). 
These results clearly illustrate the fast decay of $g_{mn,\k\a,\k'\a'}^{(2)}(\bR_p, \bR_q, \bR_r)$ with respect to both electronic and nucleus degrees of freedom in real space, thus demonstrating the feasibility of Wannier interpolation of the second-order electron-phonon coupling matrix elements.} 

\section{Convergence of the polaron equations with the number of bands}
\label{App:band_conv}

\begin{figure}
    \centering
    \includegraphics[width=0.99\linewidth]{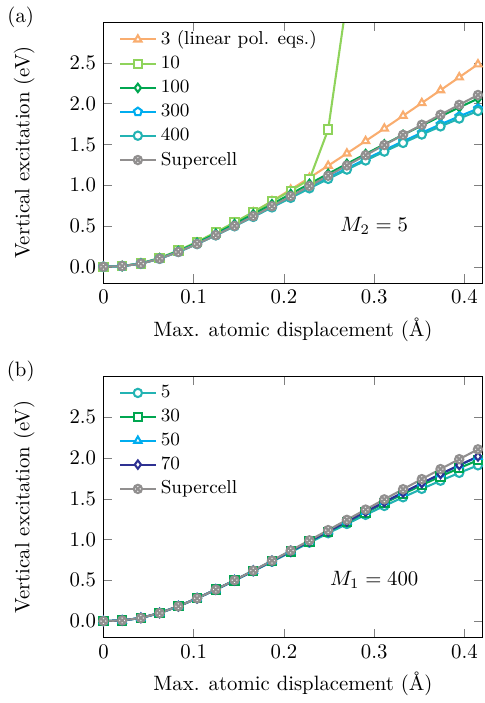}
    \caption{
    Convergence of the vertical excitation energy of the one-center, small hole polaron in LiF with respect to the number of bands included in the second-order polaron equations. (a) Calculations performed by setting $M_2=5$ in Eq.~\eqref{Eq:2nd_plrneq_1_rec}, and increasing $M_1$ from 5 to 400.
    For comparison, we also show the same quantity obtained by solving the \textit{linear} polaron equations and including only the three highest valence bands ($M_1=3$, orange).    
    (b) Same as in panel (a), but this time we set $M_1=400$ and vary $M_2$ between 5 and 70. 
    In both panels, gray lines and symbol corresponds to direct supercell calculations of the polaron.}
    \label{Fig:Conv_band}
\end{figure}

In the solution of the quadratic polaron equations, Eqs.~\eqref{Eq:2nd_plrneq_1_rec} and \eqref{Eq:2nd_plrneq_2_rec}, the sums over bands should run over all occupied and unoccupied Kohn-Sham states. In practice, to contain computational cost, we restrict these sums to $M_1$ bands for the first-order contribution and $M_2$ for the second-order contribution. Clearly the results must be tested for convergence with respect to $M_1$ and $M_2$.

To this end, we calculate the vertical excitation energy of the small hole polaron in LiF via Eq.~\eqref{Eq:2nd_plrneq_1_rec} as a function of atomic displacements, exactly as we did in Fig.~\ref{Fig:Energies}(c). We repeat these calculations by varying $M_1$ and $M_2$ and investigate the convergence using direct supercell calculations as the ground truth. 

In principle, we should use $M_1=M_2$ to remain within the same Hilbert space. In practice, however, calculating and storing second-order electron-phonon coupling matrix elements is much more demanding than for the linear coupling, therefore we investigate a hybrid strategy where $M_2 < M_1$.

In Fig.~\ref{Fig:Conv_band}(a), we set $M_2=5$ to retain all the valence bands, and we repeat calculations by taking $M_1=10$, 100, 300, and 400. This choice corresponds to including all the valence bands plus 5, 95, 295, or 395 conduction bands, respectively. 

We find that, compared to the direct supercell calculations, the choice $M_1=10$ leads to substantial overestimation of the excitation energy; furthermore, this choice yields much worse results than a calculation with only the linear electron-phonon couplings and three valence bands.
Increasing $M_1$ to 100 yields good agreement with the direct supercell calculation.
However, further increasing $M_1$ to 300 or 400 leads to a slight underestimation of the supercell result.

Next, in Fig.~\ref{Fig:Conv_band}(a) we fix $M_1=400$ and gradually increase $M_2$ from 5 to 70. In this case, the discrepancy between the polaron equations and the supercell result decreases systematically, with only marginal changes between 50 and 70 bands. Converge is thus achieved, and therefore we choose $M_1=400,\,M_2=70$ for the calculations leading to Figs.~\ref{Fig:Energies} and \ref{Fig:Hopping}.

This analysis indicates that convergence with respect to the number of bands requires careful control. When only linear couplings are included in the polaron equations, a minimal set of valence bands is sufficient to achieve reasonable accuracy for hole polarons. In contrast, when second-order couplings are incorporated, convergence requires a very large $M_1$ and a relatively large $M_2$.

We also observe partial cancellation between first- and second-order contributions. This behavior is reminiscent of the cancellation between the Fan–Migdal and Debye-Waller self-energies in band structure renormalization calculations~\cite{Ponce_Gonze_2025, Giustino_Cohen_2010}. A systematic assessment of this effect across a broader set of materials is therefore warranted.

\clearpage
\newpage
\bibliography{literature}

\end{document}